\documentclass[aps,twocolumn,superscriptaddress,floatfix]{revtex4-1}
\usepackage{graphicx,amsfonts,amssymb,amsmath,mathrsfs,hyperref,bm}

\newif\ifhyper
\hypertrue
\ifhyper
\hypersetup{
   citecolor = {red},
   colorlinks = {true}, 
   linkcolor = {blue},
   urlcolor = {blue} 
}
\fi

\def\be{\begin{equation}}
\def\ee{\end{equation}}
\def\bea{\begin{eqnarray}}
\def\eea{\end{eqnarray}}

\newcommand{\ket}[1]{|#1\rangle}

\newcommand{\expectval}[1]{\langle #1 \rangle}

\begin{document}

\title{A universal tensor network algorithm for any infinite lattice}

\author{Saeed S. Jahromi}
\email{jahromi@physics.sharif.edu}
\affiliation{Department of Physics, Sharif University of Technology, Tehran 14588-89694, Iran}
\affiliation{Donostia International Physics Center, Paseo Manuel de Lardizabal 4, E-20018 San Sebasti\'an, Spain}

\author{Rom\'an Or\'us}
\affiliation{Institute of Physics, Johannes Gutenberg University, 55099 Mainz, Germany}
\affiliation{Donostia International Physics Center, Paseo Manuel de Lardizabal 4, E-20018 San Sebasti\'an, Spain}
\affiliation{Ikerbasque Foundation for Science, Maria Diaz de Haro 3, E-48013 Bilbao, Spain}

\begin{abstract}
We present a general graph-based Projected Entangled-Pair State (gPEPS) algorithm to approximate ground states of nearest-neighbor local Hamiltonians on any lattice or graph of infinite size. By introducing the structural-matrix which codifies the details of tensor networks on any graphs in any dimension $d$, we are able to produce a code that can be essentially launched to simulate any lattice. We further introduce an optimized algorithm to compute simple tensor updates as well as expectation values and correlators with a mean-field-like effective environments. Though not being variational, this strategy allows to cope with PEPS of very large bond dimension (e.g., $D=100$), and produces remarkably accurate results in the thermodynamic limit in many situations, and specially when the correlation length is small and the connectivity of the lattice is large.  
We prove the validity of our approach by benchmarking the algorithm against known results for several models, i.e., the antiferromagnetic Heisenberg model on a chain, star and cubic lattices, the hardcore Bose-Hubbard model on square lattice, the ferromagnetic Heisenberg model in a field on the pyrochlore lattice, as well as the $3$-state quantum Potts model in field on the kagome lattice and the spin-$1$ bilinear-biquadratic Heisenberg model on the triangular lattice. We further demonstrate the performance of gPEPS by studying the quantum phase transition of the $2d$ quantum Ising model in transverse magnetic field on the square lattice, and the phase diagram of the Kitaev-Heisenberg model on the hyperhoneycomb lattice. Our results are in excellent agreement with previous studies.  

\end{abstract}

\maketitle

\section{Introduction}
\label{sec:intro}
 In recent years, tensor network (TN) states and methods \cite{Orus2014,Orus2014a} have been recognized as powerful tools in different areas of physics such as quantum information theory, condensed matter physics and, recently, even quantum gravity. From the perspective of condensed matter, TN methods are widely used to understand quantum many-body systems \cite{Cirac2009,Verstraete2008}, both theoretically and numerically. In one spatial dimension, Matrix Product States (MPS) \cite{Fannes1992,Ostlund1995} provide an efficient representation for the ground-state of $1d$ gapped local Hamiltonians based on their entanglement structure. MPS is also the variational wave function generated by the Density Matrix Renormalization Group (DMRG) \cite{White1993,White2004} and the time evolution block decimation method (TEBD) \cite{Vidal2003,Vidal2004}. Projected Entangled-Pair States (PEPS) \cite{Verstraete2004,Verstraete2006} are a generalization of MPS, and provides an ansatz for the ground-state of quantum many-body systems in higher dimensions. The infinite-size version of PEPS (iPEPS) \cite{Vidal2007,Orus2009} has also been put forward for studying the ground-state properties of $2d$ systems in the thermodynamic limit, and has been successfully applied to many different models \cite{Corboz2014a,Corboz2012a,Corboz2013,Corboz2014,Matsuda2013,OsorioIregui2014,Jahromi2018,Jahromi2018a}.

Despite its many virtues, a problem with the iPEPS algorithm is that it needs to be mostly re-programmed every time that one considers a new lattice. Long story short, the idea of iPEPS is generic, but the details of the implementation are lattice-dependent. Because of this, a common strategy is to map complex $2d$ lattices to a square lattice of tensors (e.g., via some coarse-graining), in such a way that one can recycle the square-lattice code. Dealing with the square lattice \cite{Vidal2007,Orus2009,Corboz2010a,Phien2015} indeed facilitates tensor updates and effective-environment calculations via, say, boundary MPS \cite{Vidal2007}, tensor renormalization group (TRG) \cite{Levin2007,Gu2008}, and corner transfer matrix renormalization group (CTMRG) \cite{Nishino1996,Orus2009,Corboz2010a}. The calculation of such effective environments is however costly, and in practice is done up to PEPS bond dimension $D \sim 10-20$ in the best-case scenario. Thus, although recent development in TN techniques have extended the application of iPEPS to more complicated $2d$ structures such as triangle \cite{Niesen2018,Bauer2012}, honeycomb \cite{Corboz2012,OsorioIregui2014}, Kagome \cite{Corboz2012a,Picot2016}, star \cite{Jahromi2018a} and cubic \cite{Xie2012,Orus2012} lattices, many different structures are still left behind, including important $3d$ lattices such as pyrochlore, hyperhoneycomb and diamond lattices, to name a few.

In this paper, by introducing a new and efficient standard for storing the connectivity information of a TN corresponding to a given lattice structure i.e., the {\it structure} matrix, we present a generic tensor network algorithm for the simulation of nearest-neighbor local Hamiltonians on any infinite lattice. More specifically, we develop a graph-based Projected Entangled-Pair State (gPEPS) method for any infinite lattice structure or graph in any dimension $d$, assuming translation invariance. In our implementation we use a simple update (SU) algorithm to simulate imaginary-time evolution (ITE) in order to approximate the ground-state (GS) of the system on lattices with coordination number $z$, using rank-($z+1$) tensors. On top of being generic, our approach can accurately handle large PEPS bond dimension (such as $D=100$) in the thermodynamic limit. In our approach, expectation values are estimated using a mean-field-like environment, which  provides a remarkably good approximation in many cases, specially if the correlation length is small and the coordination number $z$ is large. As benchmarks, we apply our gPEPS technique to several $2d$ and $3d$ models i.e., the antiferromagnetic Heisenberg (AFH) model on a chain, star and cubic lattices, the hardcore Bose-Hubbard (HBH) model on square lattice, the spin-$1$ bilinear-biquadratic (BLBQ) Heisenberg model on the triangular lattice, the $3$-state quantum Potts (3SQP) model in field on the kagome lattice, and the ferromagnetic Heisenberg model in field (FHF) on the pyrochlore lattice. We further challenge our technique by studying the quantum phase transition (QPT) of the transverse-field Ising model (ITF) on the square lattice and the full phase diagram of the Kitaev-Heisenberg model on the hyperhoneycomb lattice.

The paper is organized as follows: In Sec.~\ref{sec:method}, we introduce the concept of structure matrix to store the connectivity information of any TN graph and on top of that, we develop the gPEPS machinery and an efficient simple-update algorithm for approximating the ground-stet of local Hamiltonians. Further discussions regarding the calculation of expectation values with both simple and full environment, as well as relation to the Bethe and Husimi trees are provided in this section. We present our energy benchmark results for different models in Sec.~\ref{sec:benchmarck} and demonstrate the performance of gPEPS technique for studying the QPT in Sec.~\ref{sec:QPT}. Finally, Sec.~\ref{sec:conclude} is devoted to conclusion and further discussions on the advantages and drawbacks of the method.

\begin{figure}
\centerline{\includegraphics[width=\columnwidth]{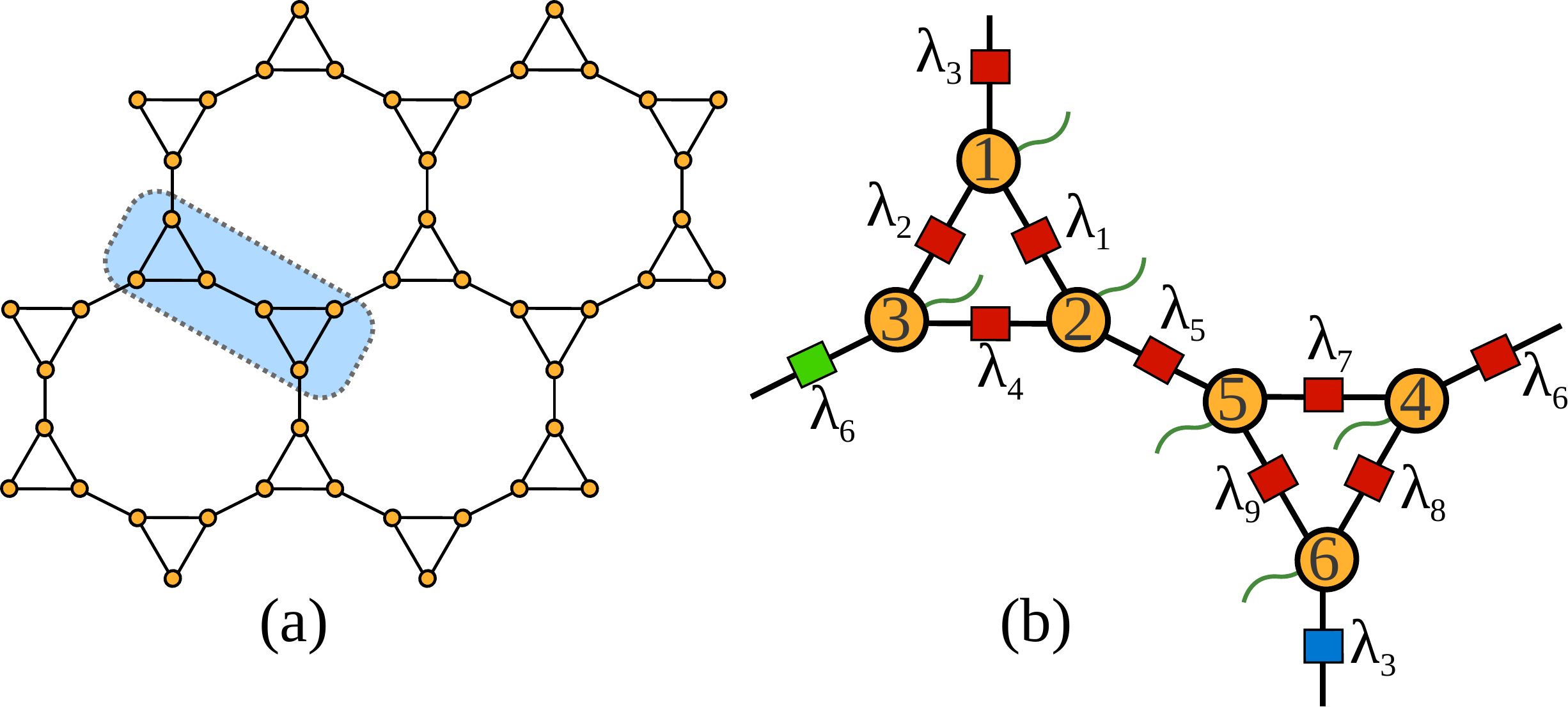}}
\caption{(Color online) (a) The $2d$ star lattice. The blue region hughlights a six-site unit cell. (b) The iPEPS TN corresponding to the star lattice unit cell.}
\label{Fig:stargraph}
\end{figure}

\section{Method}
\label{sec:method}
In this Section, we first review the basic ideas of iPEPS and how the ground state of local Hamiltonians ar represented and stored in TN language.

\subsection{${\rm g}$PEPS Basics}
\label{subsec:method}

Consider a generic infinite lattice composed of a periodically repeating unit-cell in arbitrary dimension $d$. To each vertex $i$ of the lattice, we associate a rank-$(z+1)$ iPEPS tensor $T^{s_i}_{l_1,\ldots,l_z}$, where $s$ is the physical index taking up to $p$ values for the local basis $\{\ket{\mathcal{C}}_{\mathcal{C}=1,\ldots,p}\}$, 
and $l_1,\ldots,l_z$ are virtual indices taking up to $D$ values. We also associate diagonal bond matrices $\lambda_k$ to edges $E_k$ of the lattice. In $1d$ with open boundary conditions, these $\lambda$ matrices contain the Schmidt coefficients (singular values) obtained when considering the bipartiion of one half of the system versus the other half. In two and higher dimensions, they are an approximation to the relevant degrees of freedom describing the physical system for the environment connected by the bond index.   By gluing these tensors along theirs virtual legs, we end up with a $d$-dimensional PEPS with the same structure as the original lattice. 
  
In order to approximate the GS wave function of a quantum lattice model with nearest-neighbor Hamiltonian terms $H_{i,j}$, we apply the imaginary-time evolution operator $U_{i,j}=\exp(-\delta\tau H_{i,j})$ on each edge $k$ shared between two neighboring tensors $T_i$ and $T_j$ of the PEPS, and subsequently update the $\lambda_k$ matrix as well as the $T_i$ and $T_j$ tensors. To make this as general and systematic as possible, we need extra information about the connections between neighboring tensors in the TN. More precisely, considering each local iPEPS tensor as a multidimensional array $T(p,D_1,\ldots,D_z)$, we have to know a priori which dimensions of the $T_i$, $T_j$ arrays are connected along the edge $E_k$ of the lattice so that we could update the tensors along their corresponding shared edges,each time the imaginary-time evolution operator acts on the lattice. Current state-of-the-art iPEPS algorithm typically takes care of this technical issue by mapping the $2d$ lattices to coarse-grained square structure. However, extending this strategy to any structure particularly, the $3d$ lattices, is not possible. In the next subsection, we present a generic method to resolve this problem.

\subsection{Structure Matrix}
\label{subsec:structmat}

Here we present an efficient method for storing the connectivity information of a TN corresponding to a given lattice structure. We illustrate our strategy for the example of the star lattice in $2d$ (Fig.~\ref{Fig:stargraph}-(a)). The generalization to other lattices and dimensions is straightforward (see the Appendix). Fig.~\ref{Fig:stargraph}-(b) illustrates the six-site unit cell TN of an infinite star lattice. Considering this TN as a graph in which the tensors $T_i$ correspond to graph nodes and edges $E_k$ (tensor legs) correspond to graph links, the connectivity information of the star TN is given by the so called {\it incidence} matrix \cite{Coxeter1973}:
\be
\left(\begin{tabular}{l|lllllllll}
	& $E_1$ & $E_2$ & $E_3$ & $E_4$ & $E_5$  & $E_6$ & $E_7$ & $E_8$ & $E_9$ \\
	\hline
	$T_1$     & 1 & 1 & 1 & 0 & 0 & 0 & 0 & 0 & 0  \\
	$T_2$     & 1 & 0 & 0 & 1 & 1 & 0 & 0 & 0 & 0  \\
	$T_3$     & 0 & 1 & 0 & 1 & 0 & 1 & 0 & 0 & 0  \\
	$T_4$     & 0 & 0 & 0 & 0 & 0 & 1 & 1 & 1 & 0  \\
	$T_5$     & 0 & 0 & 0 & 0 & 1 & 0 & 1 & 0 & 1  \\
	$T_6$     & 0 & 0 & 1 & 0 & 0 & 0 & 0 & 1 & 1  \\
\end{tabular}\right).
\label{Eq:incidmat}
\ee
The rows (columns) of matrix~\eqref{Eq:incidmat} correspond to tensors (edges), and the two non-zero entries in each column distinguish the two connected tensors along that edge. Although the incidence matrix already contains important data about the underlying network, crucial information regarding the corresponding bond dimensions of connected virtual indices is still missing. To fill this gap, we introduce another matrix, i.e., the {\it structure} matrix (SM) which is obtained from the incidence matrix by replacing its nonzero elements at each row by the corresponding label of the index in the tensor array: 
\be
\left(\begin{tabular}{l|lllllllll}
	& $E_1$ & $E_2$ & $E_3$ & $E_4$ & $E_5$  & $E_6$ & $E_7$ & $E_8$ & $E_9$ \\
	\hline
	$T_1$     & 2 & 3 & 4 & 0 & 0 & 0 & 0 & 0 & 0  \\
	$T_2$     & 2 & 0 & 0 & 3 & 4 & 0 & 0 & 0 & 0  \\
	$T_3$     & 0 & 2 & 0 & 3 & 0 & 4 & 0 & 0 & 0  \\
	$T_4$     & 0 & 0 & 0 & 0 & 0 & 2 & 3 & 4 & 0  \\
	$T_5$     & 0 & 0 & 0 & 0 & 2 & 0 & 3 & 0 & 4  \\
	$T_6$     & 0 & 0 & 2 & 0 & 0 & 0 & 0 & 3 & 4  \\
\end{tabular}\right).
\label{Eq:structmat}
\ee
This matrix now contains detailed information about the PEPS for the star lattice of Fig.~\ref{Fig:stargraph}-(b) and the connectivity information of two neighboring tensor along their shared edges are stored in the columns of the SM. For example, according to the second column of SM~\eqref{Eq:structmat}, the edge $E_2$ connects the bond matrix $\lambda_2$ and the dimensions $3$ and $2$ of tensors $T_1$ and $T_3$, respectively. Thanks to this information, the algorithm can automatically recognize the links and the tensors where two-body gates are applied, and implement a simple update. This is done by looping over the columns of the SM and systematically updating the iPEPS tensors along their corresponding edges, which can now be done automatically and regardless of the underlying lattice. 

Let us further remark that the SM formalism that we just introduced can also be used for simulation of systems with global symmetries, such as $U(1)$ and $SU(2)$ \cite{Singh2011,Singh2013,Bauer2011}. In this setting, edges in the graph may be directed which can be easily handled by adding a sign: outgoing (incoming) links can be distinguished in the SM with positive (negative) non-zero elements. 

Last but not least, the non-zero elements of the SM~\eqref{Eq:structmat} at each row start from $2$ which is due to the fact that the first dimension of tensors $T_i$ in our notation corresponds to the physical bonds and play no role in the connectivity of the underlying TN. One can therefore use other desired convention for labeling the virtual dimensions or use composite numbers to encode extra information in each row and column of the SM.  

\begin{figure*}
	\centerline{\includegraphics[width=17cm]{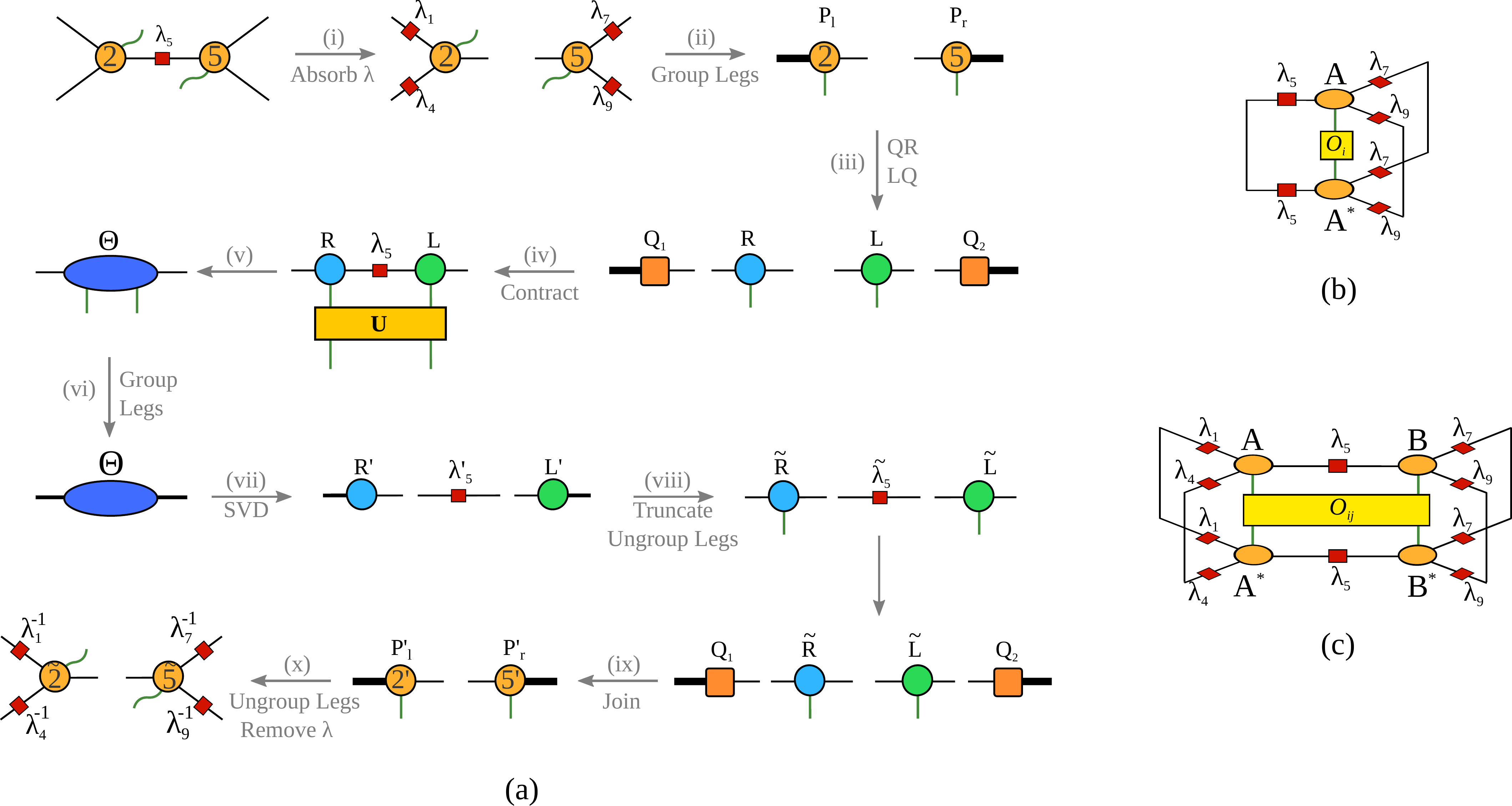}}
	\caption{(Color online) (a) Graphical representation of the SU optimization, used in the gPEPS algorithm. (b) One-site and (c) two-site expectation values, as computed with a mean-field environment, in the gPEPS scheme.}
	\label{Fig:algorithm}
\end{figure*}

\subsection{Simple Update for gPEPS}
\label{subsec:su}
 In our scheme, we approximate the ground state of a system by means of imaginary-time evolution and the simple update \cite{Corboz2010} generalized for arbitrary graphs. This method is particularly suitable for our needs, since it does not rely on an effective environment approximation (such as the full and fast-full updates \cite{Phien2015}), and is therefore implemented very similarly regardless of the lattice.

Let us now review the basics of the simple update. The ground state of a given Hamiltonian $H$, can be obtained by evolving an initial state $\ket{\Psi_0}$ in imaginary-time $\tau$ as described by
\be
\ket{\Psi_{\rm GS}}=\underset{\tau \to \infty}{\lim} \frac{e^{-\tau H} \ket{\Psi_0}}{||e^{-\tau H} \ket{\Psi_0}||}. 
\ee
When the Hamiltonian is a translationally invariant sum of nearest-neighbour terms, $H=\sum_{\langle i,j\rangle}H_{i,j}$, one can approximate the ITE operator for infinitesimal time steps $\delta\tau$ by applying a {\it Suzuki-Trotter} decomposition, i.e.,  
\be 
e^{-\delta \tau H}  \approx \prod_{\langle i,j\rangle}U_{i,j} = \prod_{\langle i,j\rangle} e^{-\delta\tau H_{i,j}}.
\ee
The GS of the system is then evaluated by iteratively applying $U_{i,j}$ on every shared link of the two neighboring tensors $T_i,T_j$ and updating the tensors along the corresponding links. In this scheme, the update changes only the tensors along the link where a given gate is acting. Therefore, one can update lower-rank sub-tensors related to them and substantially reduce the computational cost of the algorithm \cite{Phien2015}, thus allowing to achieve larger bond dimension $D$. 

Let us briefly revisit how the SU proceeds for the sub-tensors, in the context of gPEPS. Given a tensor network and its corresponding structure matrix, the SU consists of the following iterative main steps:
\begin{enumerate}
	\item Do for all edges $E_k$, $k\in[1,N_{Edge}]$ (columns of SM matrix) 
	\begin{enumerate}
		\item Find tensors $T_i,T_j$ and their corresponding dimensions connected along edge $E_k$. 
		\item Absorb bond matrices $\lambda_{m}$ to all virtual legs $m\neq k$ of $T_i,T_j$ tensors. 
		\item Group all virtual legs $m\neq k$ to form $P_l$, $P_r$ MPS tensors. 
		\item QR/LQ decompose $P_l$, $P_r$ to obtain $Q_1$,$R$ and $L$, $Q_2$ sub-tensors, respectively \cite{Phien2015}.
		\item Contract the ITE gate $U_{i,j}$, with $R$, $L$ and $\lambda_k$ to form $\Theta$ tensor.
		\item Obtain $\tilde{R}$, $\tilde{L}$, $\tilde{\lambda}_k$ tensors by applying an SVD to $\Theta$ and truncating the tensors by keeping the $D$ largest singular values (similar to $1d$ infinite TEBD \cite{Vidal2007a,Orus2008}).
		\item Glue back the $\tilde{R}$, $\tilde{L}$, sub-tensors to $Q_1$, $Q_2$, respectively, to form updated tensors $P'_l$, $P'_r$.
		\item Reshape back the $P'_l$, $P'_r$ to the original rank-$(z+1)$ tensors $T'_i,T'_j$.
		\item Remove bond matrices $\lambda_{m}$ from virtual legs $m\neq k$ to obtain the updated tensors $\tilde{T}_i$ and $\tilde{T}_j$.
	\end{enumerate}
\end{enumerate}
Fig.~\ref{Fig:algorithm}-(a) shows all these steps graphically. This process is then iterated until a convergence criteria is met. 

In order to have an efficient and universal algorithm applicable to any infinite lattice, the following remarks are in order: (i) In steps (b), (c), (g) and (h) one can locate the lambda matrices corresponding to each leg of a tensor from rows of the SM. For example, according to row three of the SM \eqref{Eq:structmat}, $\lambda_{2}$, $\lambda_{4}$ and $\lambda_{6}$ are connected to dimensions (legs) two, three and four of tensor $T_3$, respectively. One can therefore design  clever functions for absorbing (removing) $\lambda$ matrices to (from) each tensor legs as well as for grouping (un-grouping) the non-updating tensor legs by using the information stored in each row of the SM. (ii) In our SU optimization, we perform the ITE iteration starting from $\delta\tau=10^{-1}$ and gradually decrease it to $10^{-5}$ after iterating $4000$ times for each $\delta\tau$. We further check the convergence of the algorithm in each (or every $100$) step by calculating the energy and comparing it to a tolerance of the order $\epsilon=10^{-16}$. (iii) Furthermore, one can increase the stability of the SU algorithm by applying the gauge-fixing introduced in Appendix~\ref{appx:gauge}.

Let us further note that the computational cost of the SU scales as $O(pD^z)$, and evidently depends on the coordination number of the underlying lattice. Henceforth, the maximum achievable bond dimension $D$ is lattice dependent and is larger for structures with less coordination number, though structures with large $z$ usually need low $D$ because of entanglement monogamy. For example, in the case of star lattice with $z=3$, we managed to reach convergence for $D=100$ on a core$i7$ PC (with four threads) in $16$ hours. This time is quickly decreased on HPC clusters, where also larger bond dimension could be reached. 

\subsection{Expectation values and Correlators}
\label{subsec:expectval}

Once the tensors approximating a GS are found, they can be used to estimate expectation values of local operators such as local order parameters and two-point correlators. The usual procedure in iPEPS is to evaluate the effective environment surrounding some local tensors, which can be done by methods such as TRG, CTMRG, etc. These methods, however, are not easily adapted to arbitrary lattices in a systematic way. Because of this, in gPEPS we consider a simpler approach which is applicable to any graph. In this approach we use the bond matrices $\lambda$ \cite{Picot2015} (calculated during the SU optimization) in the same spirit as in one-dimensional systems \cite{Vidal2007a,Orus2008}, i.e., we close the bond indices with the $\lambda$ matrices, which is exact in one dimension, and corresponds to a mean-field approximation of the effective environment in higher dimensions. A diagrammatic representation of one- and two-site expectation values in this scheme is shown in Fig.~\ref{Fig:algorithm}-(b),(c). Similar approach has also been used in Ref.~\cite{Ran2012,Ran2013,Ran2016,Ran2017,Ran2017a,Liao2016,Picot2015,Picot2016}. Extension to other multi-site operators and correlation functions is straightforward. 

Some remarks are in order. First, due to larger bond dimension $D$ which is handled in the gPEPS algorithm, $\lambda$ matrices provide a better approximation to the environment of local tensors compared to conventional SU algorithms. Second, this scheme can be applied systematically, regardless of the underlying lattice. Third, we expect this scheme to work well in higher dimensions whenever the correlation length is small and the connectivity is large. And fourth, for $1d$ graphs, the gPEPS algorithm is exactly equivalent to the iTEBD algorithm and bond matrices satisfy the canonical forms \cite{Vidal2007a,Orus2008}, whereas in higher dimensions it provides an  approximation to expectation values which, though not being variational, may be remarkably accurate. 

%
%

\section{Energy Benchmark Results}
\label{sec:benchmarck}

\begin{table}[!t]
	\begin{center}
		\caption{gPEPS benchmark results for the GS energy per-site of several lattice models. Simulation details can be found in the supplementary material.}
		\label{tab1}
		\begin{ruledtabular}
			\begin{tabular}{l*{4}{c}r}
				Model             & Lattice & gPEPS & Previous Studies \\
				\hline
				AFH & Chain & -0.44304 & -0.44315 \cite{White1993}  \\				
				AFH & Star & -0.37523 & -0.37523 \cite{Jahromi2018a}  \\
				AFH & Cubic & -0.89253 &  -0.904 \cite{Ran2017a}  \\
				HBH & Square & -0.30258 & -0.30232 \cite{Jordan2009} \\
				FHF & Pyrochlore & -0.80000 & -0.80000   \\
				3SQP & Kagome & -4.00074 & ---   \\				
				BLBQ  & Triangular & 2.95252 & 2.95254 \cite{Niesen2018} \\
			\end{tabular}
		\end{ruledtabular}
	\end{center}
\end{table}

We benchmarked the gPEPS algorithm for several quantum lattice models, namely, the spin-$1/2$ AFH model on chain, star and cubic lattices, the HBH model on square lattice, spin-$1/2$ FHF model on pyrochlore lattice, as well as the 3SQP model in field on kagome and the spin-$1$ BLBQ Heisenberg model on the triangular lattices. Our results for the GS energy per-site of these models are summarized and benchmarked against previous studies (when it was available) in Table~\ref{tab1}, where one can clearly see the excellent agreement between our results and previous findings. Detailed discussion about each model is presented in the following.

\subsection{Antiferromagnetic Heisenberg model on $1d$ chain}
\label{subsec:simchain}

\begin{figure}[t]
	\centerline{\includegraphics[width=1.1\columnwidth]{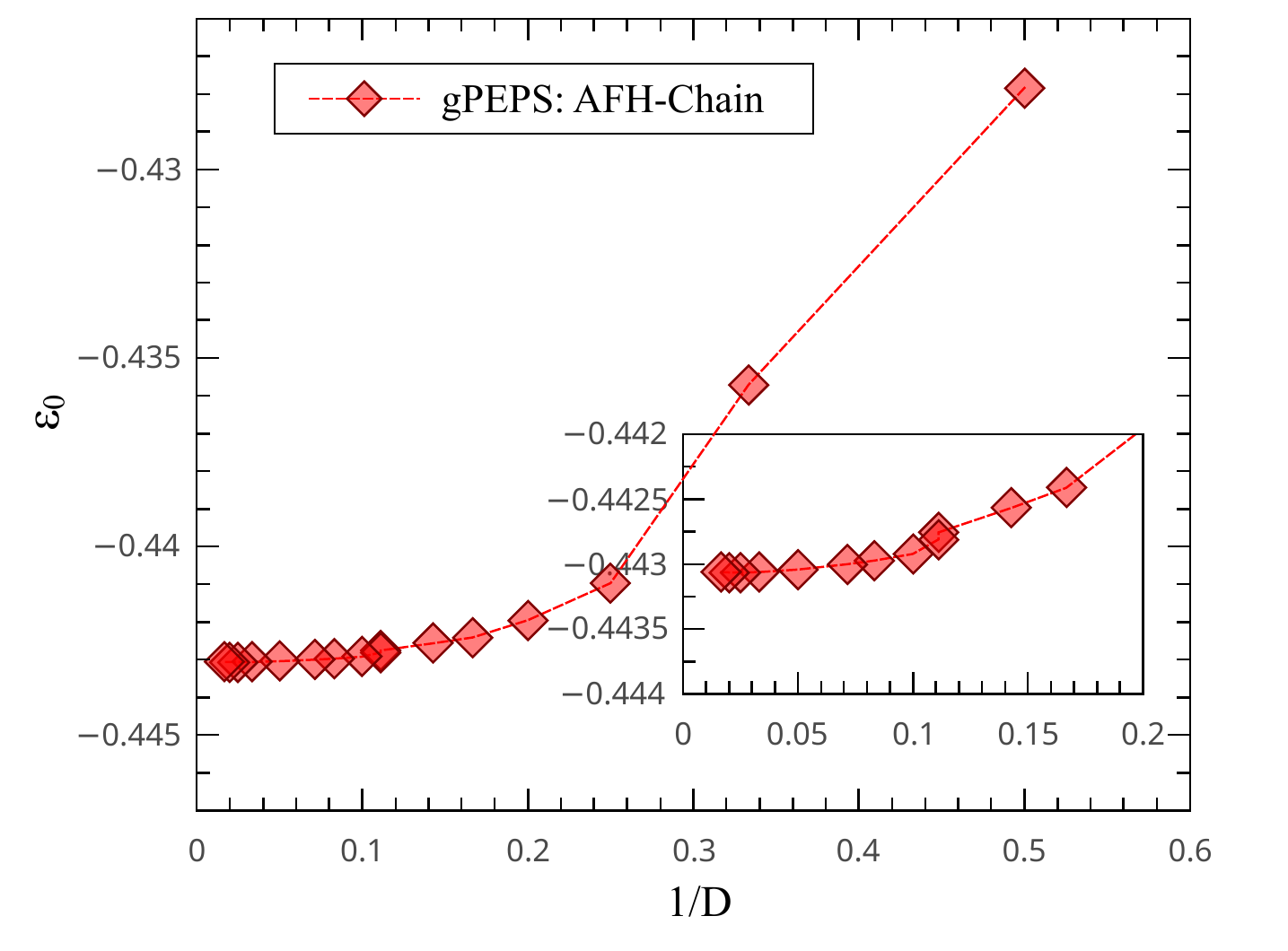}}
	\caption{(Color online) Scaling of the gPEPS ground-state energy per-site, $\varepsilon_0$, with respect to inverse bond dimension $D$ for the AFH model on $1d$ chain up to $D_{\rm Max}=60$. The inset shows the zooming for large bond dimensions.}
	\label{Fig:e0-AFH-Chain}
\end{figure}

As the first example of a lattice model, we calculate the GS energy of a $1d$ model, i.e., the spin-$1/2$ antiferromagnetic Heisenberg model on a chain. The Hamiltonian of the AFH model is given by
\be
\label{H-AFH}
H_{{\rm AFH}}=J \sum_{\langle i j \rangle} \mathbf{S}_i \cdot \mathbf{S}_j,
\ee
where the sum runs over the nearest-neighbor sites $i,j$ of the lattice and $\mathbf{S}_i$ is the ordinary spin operator at site $i$. Here we consider the antiferromagnetic Heisenberg coupling $J=1$. In order to evaluate the GS of the AFH model on a chain, we consider an infinite chain with a transitionally invariant two-site unit-cell (Fig.~\ref{Fig:chain}-(a)) and associate a rank-3 tensor to each vertices of the chain. Fig.~\ref{Fig:chain}-(b) illustrates the labelling on tensors which, corresponds to graph nodes, in the unit-cell. The corresponding SM of the chain is further given in Appendix~\ref{appx:SMChain}

Using this SM along with the simple update introduced in previous section, we evaluated the GS energy per-site, $\varepsilon_0$, of the AFH model on chain for different values of bond dimension $D$. Fig.~\ref{Fig:e0-AFH-Chain}, demonstrates the scaling of energy versus inverse bond dimension $D$ for the AFH model on $1d$ chain up to $D_{\rm Max}=60$. As one can see, there is a very good convergence for energies, particularly fore large $D$s (see also the inset of the figure). The lowest energy we obtained from gPEPS method is $\varepsilon_0=-0.44304$ which is in excellent agreement with previous density matrix renormalization group (DMRG) result, $\varepsilon_0^{DMRG}=-0.44315$, of Ref.~\cite{White1993}

As we pointed out previously, the gPEPS in $1d$ is fully equivalent to the infinite time-evolution block decimation (iTEBD) method and therefore one should obtain the exact same energy from a standard iTEBD algorithm.

\subsection{Antiferromagnetic Heisenberg model on $2d$ star lattice}
\label{subsec:simstar}

\begin{figure}[t]
	\centerline{\includegraphics[width=1.1\columnwidth]{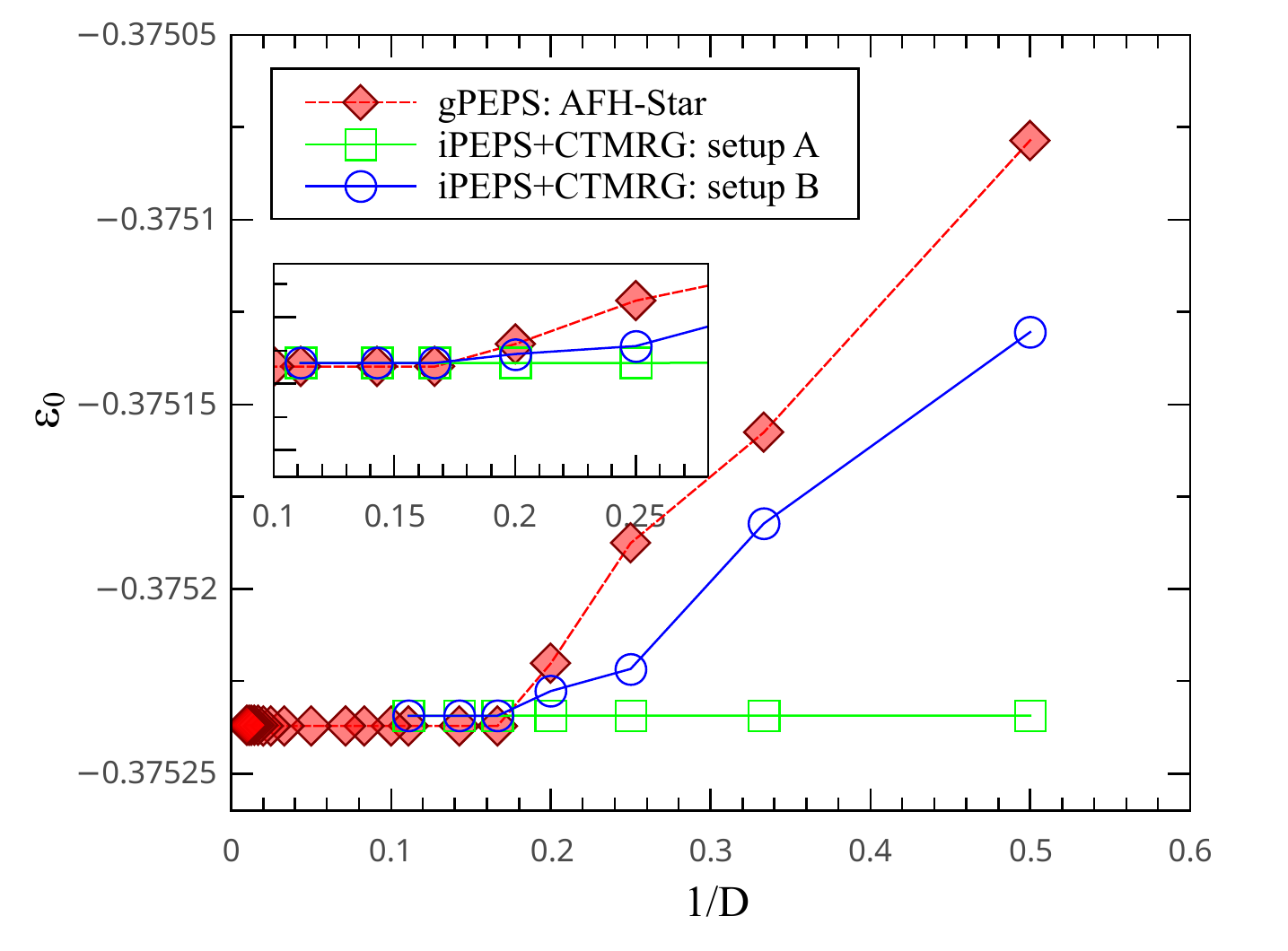}}
	\caption{(Color online) Scaling of the gPEPS ground-state energy per-site, $\varepsilon_0$, with respect to inverse bond dimension $D$ for the AFH model on $2d$ star lattice for $J_e=1, J_t=0.05$ up to $D_{\rm Max}=100$. Details of setups A, B are provided in \cite{Jahromi2018}. The inset further shows the zooming for large bond dimensions.}
	\label{Fig:e0-AFH-Star}
\end{figure}

As the second benchmark, we use the gPEPS method to calculate the GS energy of the AFH model on the star lattice. The Hamiltonian of the AFH model on the star lattice reads \cite{Jahromi2018}
\be
\label{H-AFHS}
H_{{\rm AFHS}}=J_e \sum_{\langle i j \rangle \in e} \mathbf{S}_i \cdot \mathbf{S}_j + J_t \sum_{\langle i j \rangle \in t} \mathbf{S}_i \cdot \mathbf{S}_j,
\ee
where the first sum runs over the nearest-neighbour sites on the expanding links connecting the triangles of the lattice and the second sum runs over nearest-neighbour sites on the triangles. The SM of the star lattice for a six-site unit-cell is already provided in Eq.~\eqref{Eq:structmat}.

Using \eqref{Eq:structmat}, we calculated the $\varepsilon_0$ for the AFH model on the star lattice for $J_e=1, J_t=0.05$ up to $D_{\rm Max}=100$. Fig.~\ref{Fig:e0-AFH-Star} depict the scaling of GS energy per-site for inverse of different bond dimensions. The very good convergence of energies, as well as the unprecedented large bond dimension $D_{\rm Max}=100$, definitely confirms the efficiency and power of the gPEPS technique for simulation of strongly correlated quantum many-body Hamiltonians. 

Let us further note the our gPEPS energy, $\varepsilon_0=-0.37523$, is in exact agreement with previous iPEPS study of the AFH model on the star lattice \cite{Jahromi2018}.

\subsection{Antiferromagnetic Heisenberg model on $3d$ cubic lattice}
\label{subsec:simcubic}

\begin{figure}[t]
	\centerline{\includegraphics[width=1.1\columnwidth]{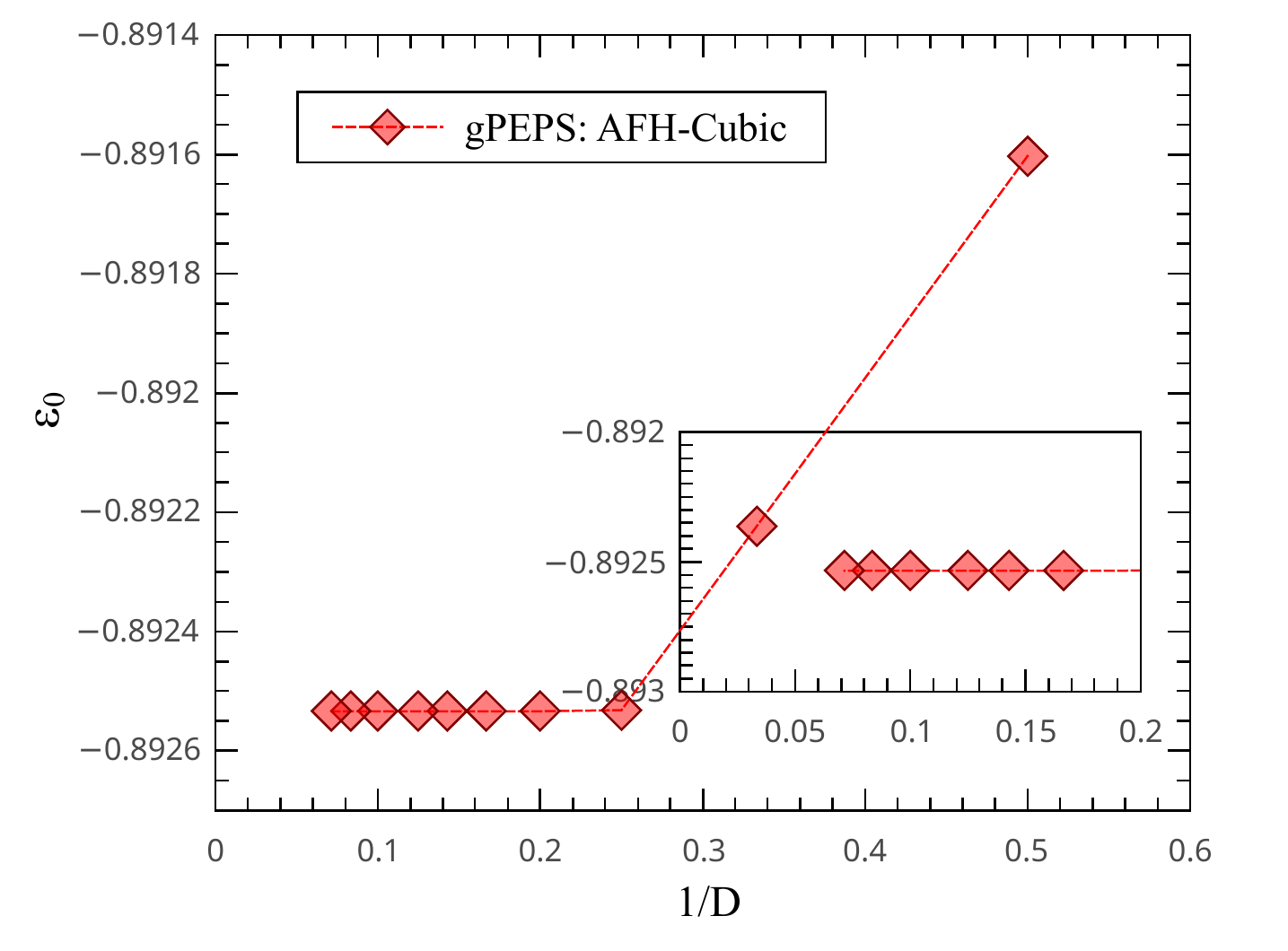}}
	\caption{(Color online) Scaling of the gPEPS ground-state energy per-site, $\varepsilon_0$, with respect to inverse bond dimension $D$ for the AFH model on $3d$ cubic lattice up to $D_{\rm Max}=14$. The inset shows the zooming for large bond dimensions.}
	\label{Fig:e0-AFH-Cubic}
\end{figure}

In order to challenge the power of gPEPS technique for $3d$ lattices, we apply it to the AFH model on the simple cubic lattice. Fig.~\ref{Fig:cubic} depicts an eight-site unit-cell of the cubic lattice and the corresponding labeling of vertices. The corresponding SM matrix is further given in Appendix~\ref{appx:SMcubic}.

Using Hamiltonian \eqref{H-AFH} and structure matrix \eqref{Eq:smcubic}, we calculated the GS energy of the AFH model on the simple cubic lattice for different bond dimensions. Fig.~\ref{Fig:e0-AFH-Cubic} shows the scaling of energy versus inverse bond dimension up to $D_{\rm Max}=14$ on the cubic lattice. The results show a very good convergence of the gPEPS energies to $\varepsilon_0=-0.89253$ which is in   close agreement with the results of Ref.~\cite{Ran2017a} with $\varepsilon_0=-0.904$. Our findings once again confirms how the idea of SM can simplify the implementation of TN methods to $3d$ lattice models.

\subsection{Hardcore Bose-Hubbard model on $2d$ square lattice}
\label{subsec:simsquare}

\begin{figure}[t]
	\centerline{\includegraphics[width=1.1\columnwidth]{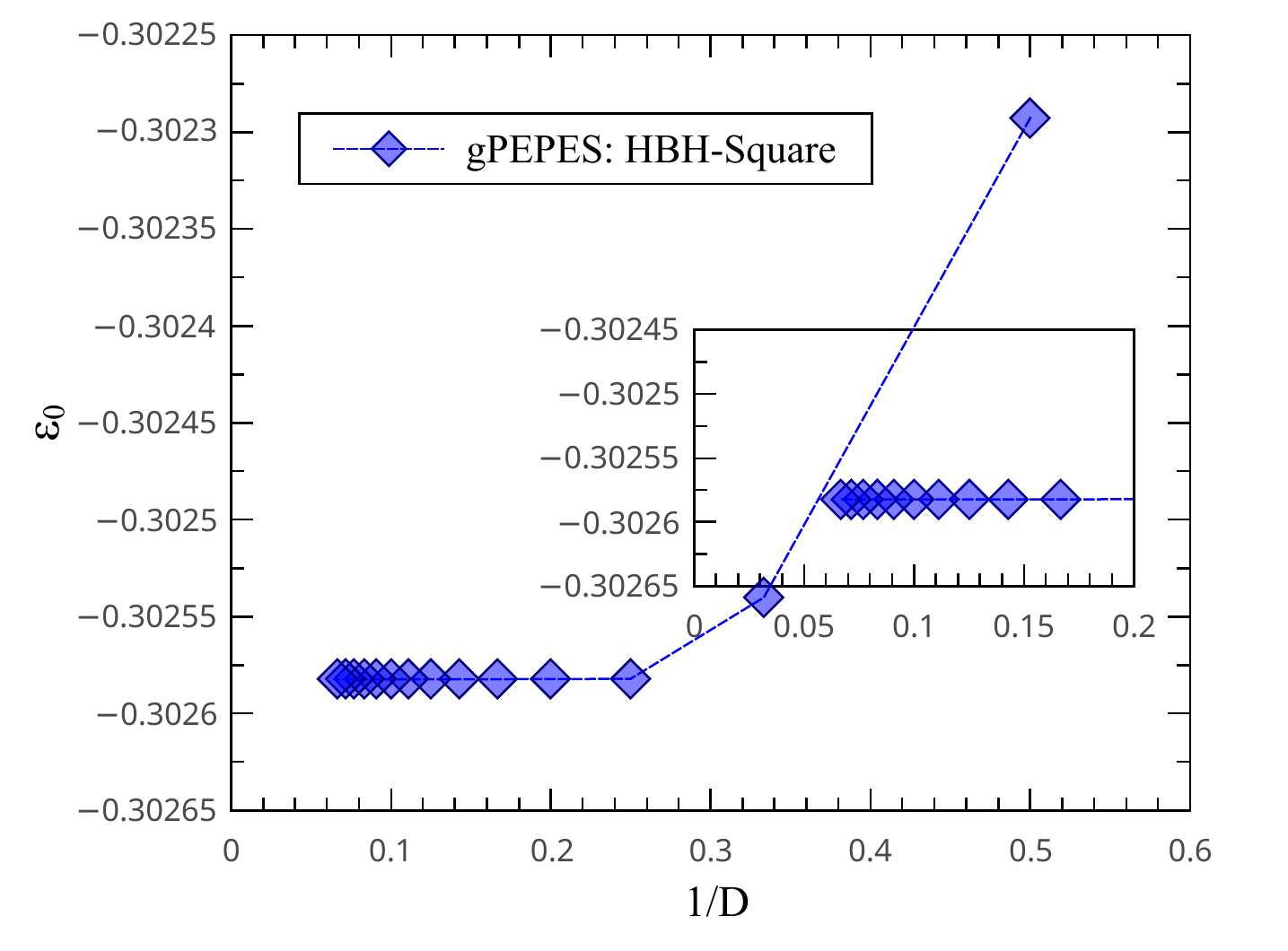}}
	\caption{(Color online) Scaling of the gPEPS ground-state energy per-site, $\varepsilon_0$, with respect to inverse bond dimension $D$ for the HBH model on $2d$ square lattice for $\mu=-2$ up to $D_{\rm Max}=14$. The inset shows the zooming for large bond dimensions.}
	\label{Fig:e0-HBH-Square}
\end{figure}

In this subsection we test our gPEPS algorithm for another lattice model, i.e., the hardcore Bose-Hubbard model on the square lattice. Fig.~\ref{Fig:square}-(a),(b) demonstrate the square lattice and the four-site unit-cell that we used for our simulation. Hamiltonian of the HBH model further reads
\be
\label{H-HBH}
H_{{\rm HBH}}=-J \sum_{\langle i j \rangle} (a^{\dagger}_i a_j+a^{\dagger}_j a_i)-\mu \sum_i \hat{n}_i,
\ee
where the first hopping term is on the nearest-neighbor vertices of the square lattice and the second sum is an on-site chemical potential. Here we set $J=1$.$a$ and $a^{\dagger}$ are bosonic annihilation and creation operators. The SM of the square lattice which is required for the gPEPS simulation is further provided in Appendix~\ref{appx:SMsquare}.

Fig.~\ref{Fig:e0-HBH-Square} demonstrate our findings for the GS energy of the HBH model for $\mu=-2$ for different bond dimensions up to $D_{\rm Max}=14$. The convergence at large $D$s are quite good and the GS energy per-site of the system for $D=14$ is $\varepsilon_0=-0.30258$ which is even lower than  previous iPEPS results of Ref.~\cite{Jordan2009} with $\varepsilon_0^{iPEPS}=-0.30232$.

\subsection{Spin-$1$ bilinear-biquadratic Heisenberg model on $2d$ triangular lattice}
\label{subsec:simtriangular}

\begin{figure}[t]
	\centerline{\includegraphics[width=1.1\columnwidth]{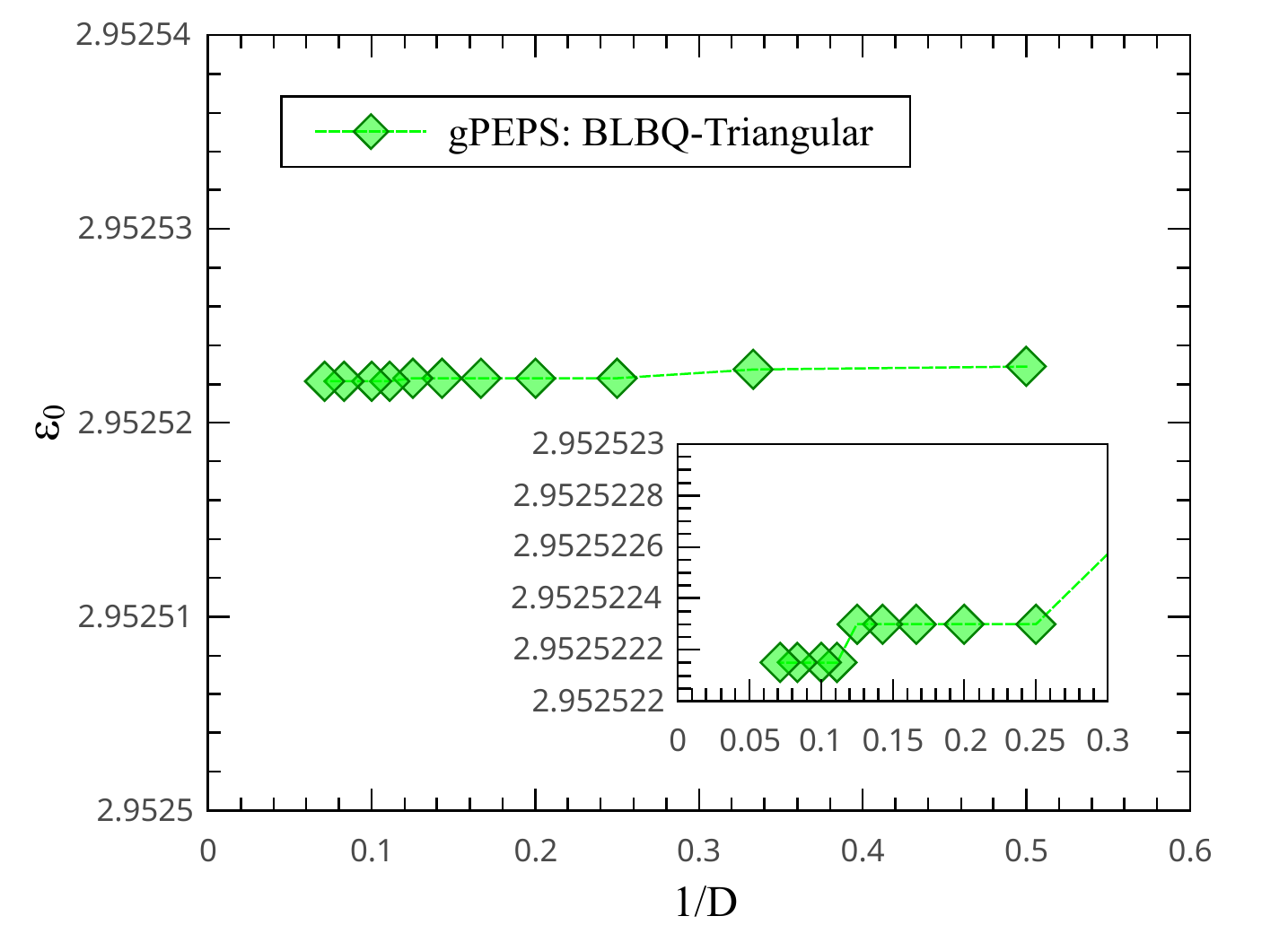}}
	\caption{(Color online) Scaling of the gPEPS ground-state energy per-site, $\varepsilon_0$, with respect to inverse bond dimension $D$ for the BLBQ model on $2d$ triangular lattice for $\theta=1.5865$ up to $D_{\rm Max}=14$. The inset shows the zooming for large bond dimensions.}
	\label{e0-BLBQ-Triangular}
\end{figure}

As another example for benchmarking the gPEPS method, we studied the spin-$1$ bilinear-biquadratic Heisenberg model on $2d$ triangular lattice (Fig.~\ref{Fig:triangular}-(a)). This model has already been studied in detail in Ref.~\cite{Niesen2018} with iPEPS method and the full phase diagram of the system has already been investigated. The iPEPS machinery for triangular lattice is performed by mapping it to square lattice with both nearest and next-nearest neighbour interactions. 

Here instead, we study the model by means of gPEPS technique on an infinite triangular lattice with nine-site unit-cell (see Fig.~\ref{Fig:triangular}-(b)). In the gPEPS framework, all of the interactions are between nearest-neighbour vertices and simulation for larger bond dimensions is also possible.

Hamiltonian of the spin-$1$ BLBQ model according to the convention of Ref.~\cite{Niesen2018} reads
\be
\label{H-BLBQ}
H_{{\rm BLBQ}}=\cos(\theta) \sum_{\langle i j \rangle} \mathbf{S}_i \cdot \mathbf{S}_j+\sin(\theta) \sum_{\langle i j \rangle} (\mathbf{S}_i \cdot \mathbf{S}_j)^2,
\ee
where both sums run on nearest-neighbours. The first sum however, is the bilinear term which is nothing but the standard Heisenberg model and the second term is the biquadratic term. 

In order to benchmark the gPEPS results with previous studies, we calculate the GS of the system for $\theta=1.5865$. This point is very close to $\theta=\frac{\pi}{2}$. However since $\theta=\frac{\pi}{2}$ is a phase boundary in the phase diagram of the BLBQ model on the triangular lattice \cite{Niesen2018}, we chose a slightly different point to evaluate the GS of the system to show how the gPEPS can converge to the true GS of the system.

Using Hamiltonian \eqref{H-BLBQ} and the SM of the triangular lattice presented in Appendix~\ref{appx:SMtriangular}, we were able to reproduce the results of Ref.~\cite{Niesen2018} with very high accuracy. Fig.~\ref{e0-BLBQ-Triangular} depicts the scaling of the gPEPS GS energy per-site, $\varepsilon_0$, with respect to inverse bond dimension $D$ for the BLBQ model for $\theta=1.5865$. As one can clearly see, the convergence of the algorithm is quite notable even at small bond dimensions and our gPEPS energy $\varepsilon_0=2.95252$ is almost the same as $\varepsilon_0^{iPEPS}=2.95253$ of the Ref.~\cite{Niesen2018}.

\begin{figure*}
	\centering
	\begin{tabular}{cc}
		\includegraphics[width=0.5\textwidth,trim={0 0 0 0},clip]{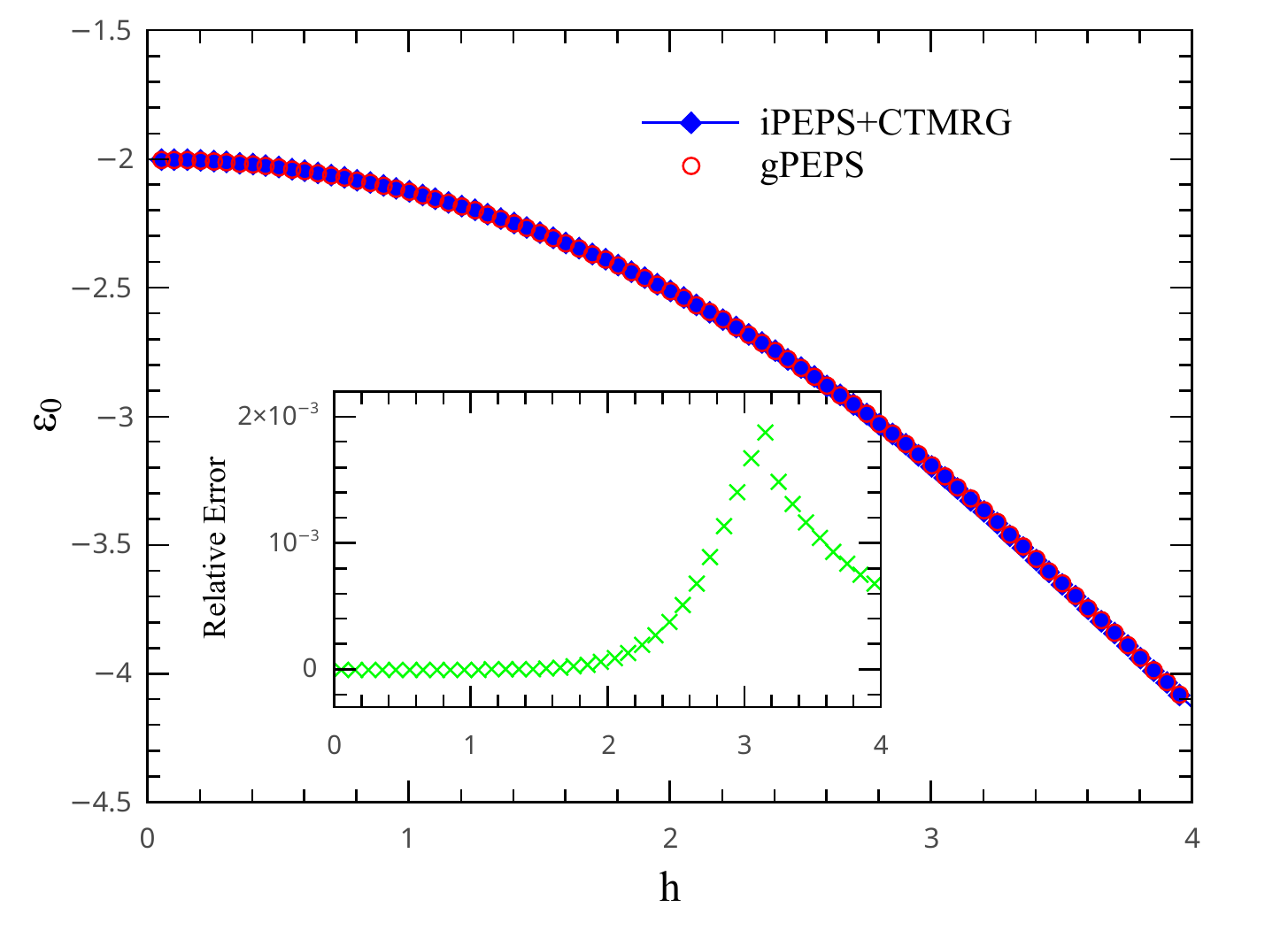} & \includegraphics[width=0.5\textwidth,trim={0 0 0 0},clip]{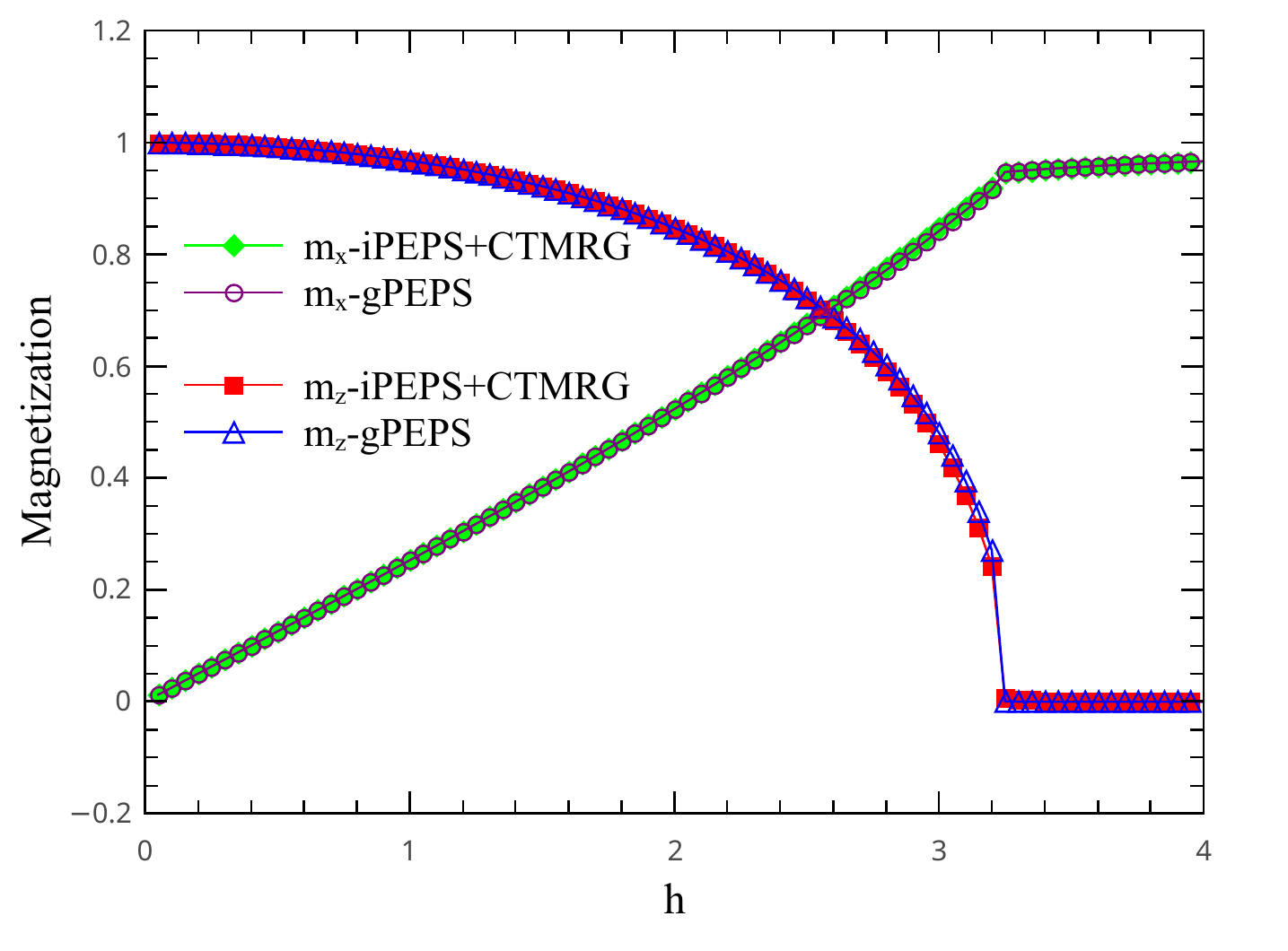} \\
		(a) & (b) 
	\end{tabular}
	\caption{(Color online) (a) The GS energy per site and magnetization $m_z$ of the ITF model with respect to field strength $h$ for the gPEPS method ($D=6$), compared with the iPEPS+CTMRG ($D=6,\chi=80$) on a $2\times2$ unit-cell. The inset in the energy plot demonstrates the gPEPS relative error with respect to the iPEPS energies.}
	\label{Fig:energymag} 
\end{figure*}

\subsection{3-State quantum Potts model in field on $2d$ kagome lattice}
\label{subsec:simkagome}

Here we present our gPEPS results for the 3-state Potts model in field on the kagome lattice which, to the best of our knowledge, is the first TN implementation of this model. Generic Hamiltonian of the q-state Potts model, also known as vector Potts model, in the presence of field reads \cite{Cobanera2011}
\be
\label{H-Potts}
H_{{\rm Potts}}=-J \sum_{\langle i j \rangle} U_i U^{\dagger}_j-\Gamma \sum_i V_i+{\rm h.c.},
\ee
where
\be
U={\rm diag}(1,\omega,\omega^2,\ldots,\omega^{q-1}), \    \  \omega=e^{\frac{2\pi i}{q}},
\ee
and
\be
V=\quad
\begin{pmatrix} 
	0 & I_{q-1} \\
	1 & 0 
\end{pmatrix}
\quad,
\ee
where $I_{q-1}$ is a $(q-1)\times(q-1)$ identity matrix. By setting $q=3$ in the above relations, Hamiltonian of the 3SQP is obtained. We then apply Hamiltonian \eqref{H-Potts} to a kagome lattice with a twelve-site unit-cell (Fig.~\ref{Fig:kagome}). The corresponding SM of the kagome unit-cell is given in Appendix~\ref{appx:SMkagome}.

We have calculated the $\varepsilon_0$ for the 3SQP model in field on the kagome lattice with the gPEPS method up to $D_{\rm Max}=30$. The GS energy of the system at this point is exact and for all bond dimensions $D$ for finite field value $\Gamma=0.1$ is equal to $\varepsilon_0=-4.00074$.

\subsection{Ferromagnetic Heisenberg model in magnetic field on $3d$ pyrochlore lattice}
\label{subsec:simpyrochlore}

In order to challenge the gPEPS algorithm with a non-trivial $3d$ lattice, we applied it to one of the most complicated structures, i.e, the pyrochlore lattice and studied the FHF model on this lattice. We stress that, to the best of our knowledge, this is the first application of TN methods to the pyrochlore lattice. 

Hamiltonian of the FHF model is given by
\be
\label{H-FHF}
H_{{\rm FHF}}=-J \sum_{\langle i j \rangle} \mathbf{S}_i \cdot \mathbf{S}_j-h \sum_i \mathbf{S}_z,
\ee
where the first sum is on nearest-neighbor sites and the second sum runs over all of the vertices of the lattice. Here we set $J=1$. We apply Hamiltonian \eqref{H-FHF} to the pyrochlore lattice (Fig.~\ref{Fig:pyrochlor}-(a)) with an eight-site unit-cell (Fig.~\ref{Fig:pyrochlor}-(b)).  The corresponding SM of the pyrochlore lattice is given in Appendix~\ref{appx:SMpyrochlore}.

The FHF model on pyrochlore lattice has an exact mean-field ground-state with energy
\be
\varepsilon_0^{\rm exact}=\frac{1}{N_s}(-\frac{N_s h}{2}-\frac{N_b J}{2}),
\ee 
which is a state with $D=1$ and thus no correlations. It is simply the state with all spins aligned in $z$-direction. In the above relation, $N_s$ is te number of lattice sites and $N_b$ is the number of nearest-neighbor bonds of pyrochlore lattice. For a translationally invariant unit-cell of the pyrochlore lattice with $8$ sites and $24$ bonds, such as the one depicted in Fig.~\ref{Fig:pyrochlor}-(b), GS energy per-site of the system for $h=0.1$ is $\varepsilon_0^{\rm exact}=-0.8$.

Our gPEPS results for the GS energy of the FHF model for $h=0.1$ is in exact agreement with the mean-field results and we obtained $\varepsilon_0=-0.80000$ for different bond dimensions $D$ up to $D_{\rm Max}=14$. This once again certifies that the gPEPS technique is a powerful universal TN method for simulation of lattice Hamiltonians on the exotic lattice structures.

\begin{figure*}
	\centerline{\includegraphics[width=18cm]{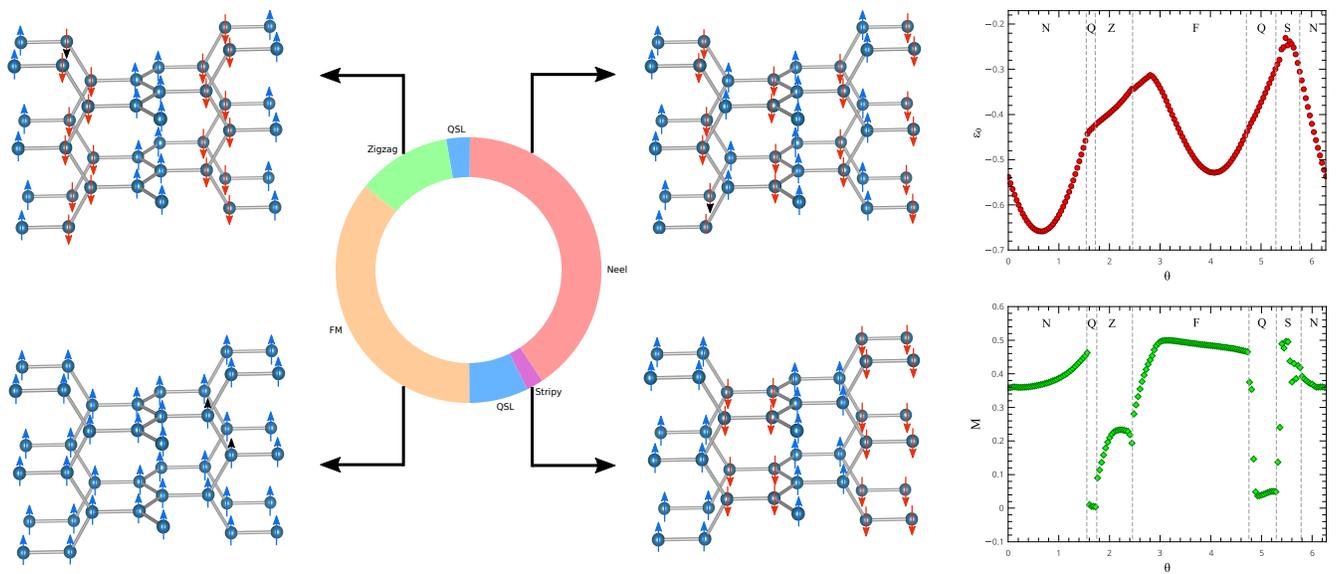}}
	\caption{(Color online) (left) Full phase diagram of the KH model on the hyperhoneycomb lattice. (right) GS energy per-site, $\varepsilon_0$, and magnetic order parameter $M$ for $\theta=[0,2\pi]$. Details of the phase diagram can be found in the main text.}
	\label{Fig:10-3-b-phase-diag}
\end{figure*}

\section{Quantum Phase Transition with ${\rm g}$PEPS}
\label{sec:QPT}

Next, we challenged the gPEPS technique for systematic study of QPT in quantum lattice models on different structures. In what follows we present our results for phase diagram of the ITF model on square lattice which is an standard benchmark model for any new algorithm and then we investigate the non-trivial phase diagram of the Kitaev-Heisenberg model on the hyperhoneycomb lattice which is one of the most complicated structures for any numerical technique. 

\subsection{Quantum Ising model in transverse magnetic field on $2d$ square lattice}
\label{subsec:simITF}

Using the gPEPS algorithm, we studied the zero-temperature phase diagram of the ITF model on a square lattice. In particular, we studied the following Hamiltonian
\be
H_{{\rm ITF}}=-J \sum_{\langle i j \rangle} \sigma_i^z \sigma_j^z-h \sum_i \sigma_i^x,
\ee
where the first sum runs over nearest-neighbor sites and the second one runs over the vertices of the square lattice. By measuring the GS energy and magnetization along $x$ and $z$-directions, we pinpointed the QPT point at $h^c\approx3.04$ which is in perfect agreement with previous studies \cite{Jordan2009,Orus2009}. Fig.~\ref{Fig:energymag} shows the GS energy per-site as well as the magnetization of the ITF model. The QPT is best captured by discontinuities in the magnetization and energy plots. 

The gPEPS relative error with respect to the iPEPS energies in the inset of Fig.~\ref{Fig:energymag}-(a) is of the order ~$10^{-16}$ everywhere except at the vicinity of the transition point which is increased to ~$10^{-3}$. This is best explained by the fact that at the critical point, the correlation length diverges and the mean-field environment does not necessarily provide the best approximation of the iPEPS environment. Nonetheless, the gPEPS still captures the QPT with very high accuracy in most cases.

\subsection{Kitaev-Heisenberg model on $3d$ hyperhoneycomb lattice}
\label{subsec:simKH}

Spin-orbit entangled Mott insulators in Iridates \cite{Khaliullin2005,Jackeli2009} can realize instances of $2d$ \cite{Chaloupka2010,Chaloupka2013} and $3d$ \cite{Modic2014,Takayama2015} arrangements of tricoordinated lattices with Kitaev interactions. In particular, it has been shown that the polymorph $\beta-Li_2IrO_3$ realize three-dimensional arrangements of the spin-orbit tangled moments which retain the hyperhoneycomb lattice \cite{Takayama2015}. In this material, the $Ir^{4+}$ ions arrange in a hyperhoneycomb structure and the combined effect of spin-orbit coupling, Coulomb interaction, and exchange geometry generates
$J_{\rm eff}=1/2$ moments subject to a combination of anisotropic Kitaev and Heisenberg interactions \cite{Khaliullin2005,Jackeli2009}. In $2d$ it has been shown that the resulting Kitaev-Heisenberg (KH) model host various phases ranging from quantum spin-liquid (QSL) to magnetically ordered phases such as ferromagnetic (FM), antiferomagnetic (AFM), stripy and zigzag on the honeycomb lattice \cite{Chaloupka2010,Chaloupka2013}. 

Recent studies based on mean-field theory \cite{Lee2014} and TN on Bethe lattice \cite{Kimchi2014} have also predicted similar phases for the  KH model on the hyperhoneycomb lattice. However a systematic study of the full phase digram of the model on the original hyperhoneycomb lattice in the thermodynamic limit is still missing. Thanks to the gPEPS technique, we were able to apply, for the first time, the TN method, directly to a translationally invariant unit-cell of the  hyperhoneycomb lattice (see Fig.~\ref{Fig:hyperhoneycomb}) and map out the phase diagram of the KH model on the full parameter space. More specifically, we applied the gPEPS to the following KH Hamiltonian
\be
\label{H-KH}
H_{{\rm KH}}=2\cos(\theta) \sum_{\alpha-link} S_i^\alpha S_j^\alpha+\sin(\theta) \sum_{\langle i j \rangle} \mathbf{S}_i \cdot \mathbf{S}_j,
\ee
where the first sum is the Kitaev term with ($\alpha=x,y,z$) and the second term is the Heisenberg interaction acting on the nearest-neighbor sites of the hyperhoneycomb lattice. 

In order to capture the phase boundaries and characterize the nature of underlying phases, we calculated the GS energy, entanglement entropy, magnetization, ground-state fidelity and two-site spin-spin correlators in the full parameter space $\theta=[0,2\pi]$. Fig.~\ref{Fig:10-3-b-phase-diag}-(left) demonstrates the phase diagram of the KH model on the hyperhoneycomb lattice. The phase diagram is composed of four magnetically ordered phases i.e., FM, AFM, zigzag and stripy and two QSL phase at the vicinity of the FM and AFM  Kitaev couplings. Orientation of spins in each magnetic phase has also been shown in the figure. detailed discussion regarding the QSL phase at the pure Kitaev points can be found in Ref.~\cite{OBrien2016}. Fig.~\ref{Fig:10-3-b-phase-diag}-(right) further show the GS energy per-site as well as the magnetic order parameter, $M=\sqrt{\expectval{S^x}^2+\expectval{S^y}^2+\expectval{S^z}^2}$. One can clearly see that $M$ precisely detect the phase boundaries and distinguishes magnetic phases from QSL phases with no local order parameter. 

We refer the interested reader to Ref.~\cite{Jahromi} for further details regarding this model and the phase diagram of KH model on other $3d$ tricoordinated lattices \cite{OBrien2016}. Let us further stress that our findings are in excellent agreement with previous studies \cite{Kimchi2014,Lee2014}


\section{Conclusions and discussion}
\label{sec:conclude}
In this paper we introduced the concept of structure matrix which encodes the connectivity information of a given tensor network and developed a generic graph-based Projected Entangled-Pair State algorithm for local Hamiltonians of quantum lattice models that can be applied to any lattice in any dimension in the thermodynamic limit. Our approach relies on the simple update algorithm for imaginary-time evolution, and a mean-field-like approximation to effective environments. Though not being variational, the scheme produces accurate results in most situations and is capable of handling large bond dimensions such as $D \sim 100$. 

We benchmarked our method with several quantum lattice models on different structures in one, two and three dimensional lattices. Our method facilitates the applicability of iPEPS algorothms to complex lattices in $2d$ and $3d$. Most importantly, it also opens the possibility to simulate quantum materials on complex crystallographic structures via tensor network methods. The gPEPS method can further be extended to deal with fermionic systems and symmetric tensor networks, as well as finite temperature. 

Let us further remark that the gPEPS ground-state tensors of all infinite $2d$ systems can additionally be contracted by using TRG, bounday MPS or CTMRG both directly or rather by grouping several adjacent tensors into a coarse-grained square lattice of block-sites in order to obtain variational energies. Unfortunately extension of these ideas to generic $3d$ structures is not straightforward. For example, the CTMRG has only been extended to simple cubic lattice and other $3d$ lattices are left behind. 
A new generic technique for contracting infinite lattices both in $2d$ and $3d$ is currently under development by our group \cite{Jahromia} which can be used as a supplement to gPEPS method for doing variational optimization with TN on any infinite graph. With this new approach we will be able to do full update within the gPEPS framework.

It is worth noting that extension of TN methods to generic lattices can alternatively be done by using Husimi lattices \cite{Liao2016} which are obtained from a Bethe lattice in which every vertex is replaced by a $p$-polygon \cite{Ostilli2012,DeMiranda-Neto1993}. Nevertheless, one must note that the physics obtained on the Husimi lattice might be different from the one on the original lattice. This is mainly due to the slight differences between a lattice and its Husimi counterpart. For example, a Husimi lattice might not create the same closed loop structure as the original lattice. This is important particularly for those models in which closed loops of the lattice play key roles in the physics of the system. For example it is already known that the closed loops of the Kitaev model act as Integrals of motion which carries zero fluxes in the system \cite{Kitaev2006}.  This once again shows the significance of gPEPS in studying generic infinite lattice with TN methods.

As last remark, let us point out that although gPEPS technique produces reliable and accurate results for many quantum lattice models in different dimensions, applications of the method to frustrated system should be handled with care. Due to the longer range of correlations which might exist in the GS of frustrated systems such as some spin-liquid states, the role of environment around local GS tensors becomes very important, and the bond matrices $\lambda$ which are used in gPEPS method as mean-field environment for calculation of the expectation values might not provide the best approximation to the environment. One might therefore obtain higher or unexpectedly lower values for the GS energies of the system and expectation values. For example the gPEPS method fails in producing accurate results for the AFH model on the kagome lattice and the best TN results so far, belongs to the projected entangled-simplex state \cite{Xie2014,Liao2017}. It is therefore advised that the gPEPS energies for frustrated system be benchmarked against other methods to make sure the correct results are obtained.

\acknowledgements
S.S.J. acknowledges the support from Iran Science Elites Federation (ISEF). The gPEPS calculations were performed on the HPC cluster at Sharif University of Technology.


%

\newpage
\onecolumngrid
\appendix

\section{Structure Matrix for various lattice structures}
\label{appx:structmat}

In This appendix, we present structure matrix of various widely used $2d$ and $3d$ lattices. The main strategy to construct the SM corresponding to a given infinite lattice with translational invariance is to first define a unit-cell of the lattice with desired number of vertices and periodic boundary condition and then constructing the incidence matrix (IM) of the unit-cell. The SM can then be obtained straightforwardly from the IM of the lattice. The IM of arbitrary graphs can be obtained by using efficient graph libraries of Matlab, Python, Mathematica or other desired languages. 

\subsection{$1d$ chain}
\label{appx:SMChain}

\begin{figure}[h]
	\centerline{\includegraphics[width=8cm]{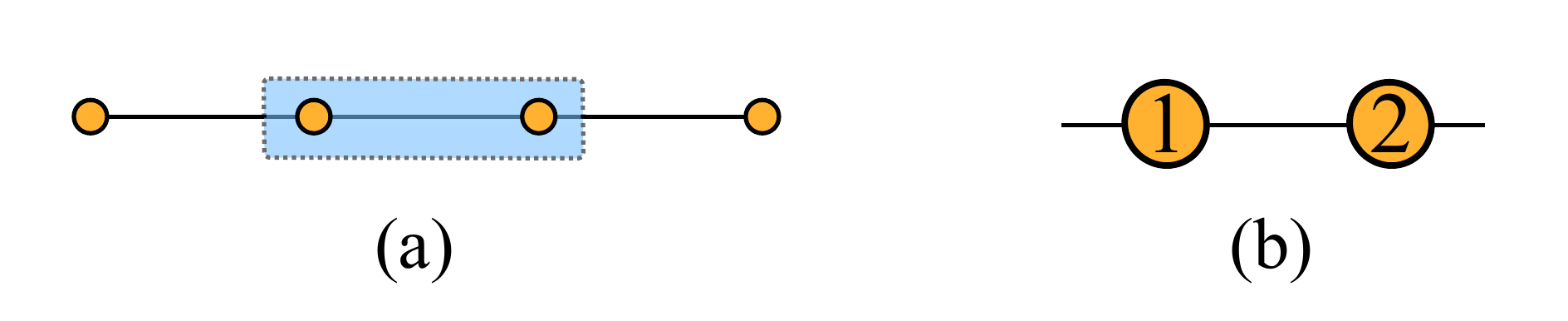}}
	\caption{(Color online) (a) The infinite $1d$ spin chain with a two-site unit-cell (blue region). (b) Labeling of vertices (graph nodes) in the unit-cell.}
	\label{Fig:chain}
\end{figure}

Eq.\eqref{Eq:smchain} corresponds to the SM of an infinite $1d$ spin chain with a two-site unit-cell (See Fig.~\ref{Fig:chain}).

\be
SM_{\rm chain}=\left(\begin{tabular}{l|ll}
	& $E_1$ & $E_2$  \\
	\hline
	$T_1$     & 2 & 3   \\
	$T_2$     & 2 & 3   \\
	
\end{tabular}\right).
\label{Eq:smchain}
\ee

\subsection{$2d$ square lattice}
\label{appx:SMsquare}

\begin{figure}[h]
	\centerline{\includegraphics[width=7cm]{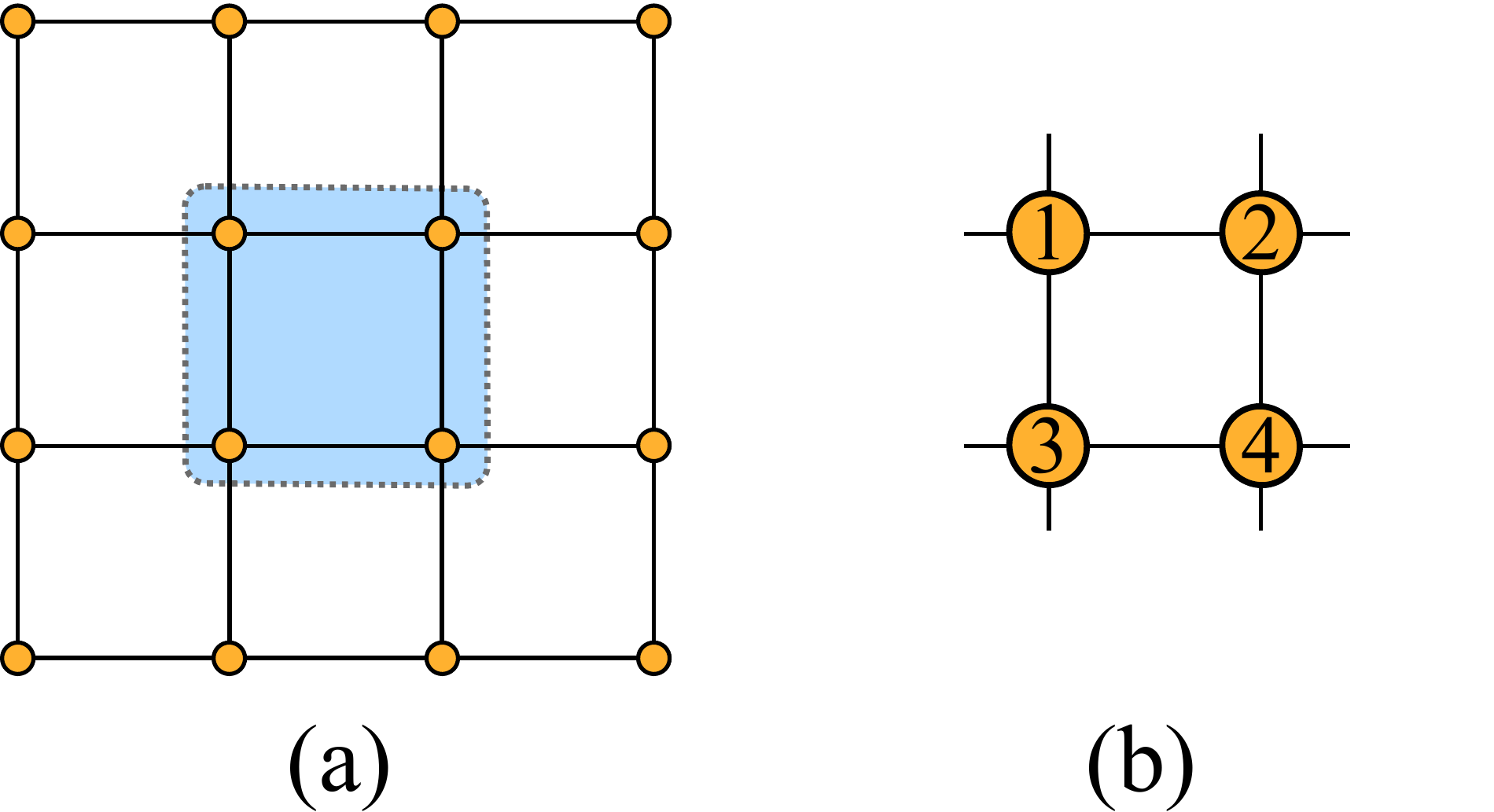}}
	\caption{(Color online) (a) The infinite $2d$ square lattice with a 4-site unit-cell (blue region). (b) Labeling of vertices (graph nodes) in the unit-cell.}
	\label{Fig:square}
\end{figure}

Eq.\eqref{Eq:smsquare} corresponds to the SM of an infinite $2d$ square lattice with a 4-site unit-cell (See Fig.~\ref{Fig:square}).

\be
SM_{\rm square}=\left(\begin{tabular}{l|llllllll}
	& $E_1$ & $E_2$ & $E_3$ & $E_4$ & $E_5$  & $E_6$ & $E_7$ & $E_8$ \\
	\hline
	$T_1$     & 2 & 3 & 4 & 5 & 0 & 0 & 0 & 0  \\
	$T_2$     & 2 & 3 & 0 & 0 & 4 & 5 & 0 & 0  \\
	$T_3$     & 0 & 0 & 2 & 3 & 0 & 0 & 4 & 5  \\
	$T_4$     & 0 & 0 & 0 & 0 & 2 & 3 & 4 & 5  \\
\end{tabular}\right).
\label{Eq:smsquare}
\ee

\subsection{$2d$ triangular lattice}
\label{appx:SMtriangular}

\begin{figure}[h]
	\centerline{\includegraphics[width=10cm]{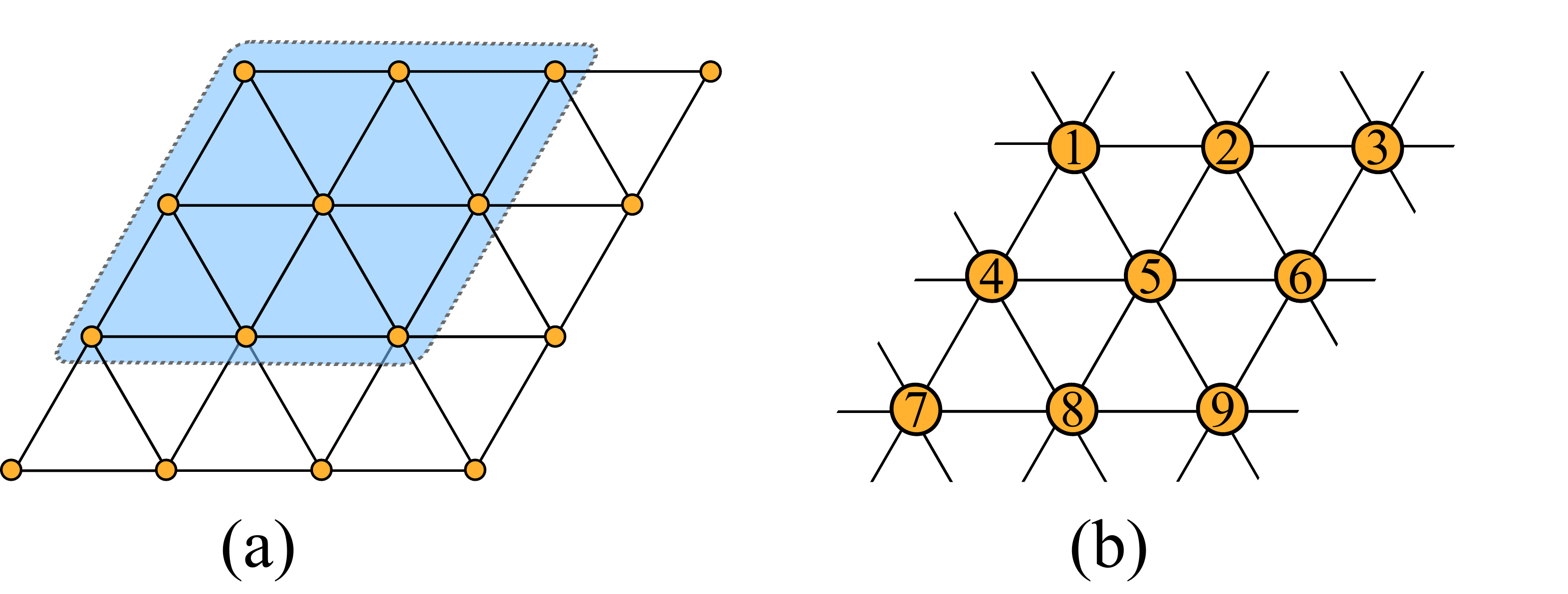}}
	\caption{(Color online) (a) The infinite $2d$ triangular lattice with a 9-site unit-cell (blue region). (b) Labelling of vertices (graph nodes) in the unit-cell.}
	\label{Fig:triangular}
\end{figure}

Eq.\eqref{Eq:smblbq} corresponds to the SM of an infinite $2d$ triangular lattice with a 9-site unit-cell (See Fig.~\ref{Fig:triangular}).

\be
SM_{\rm triang}= {\scriptsize \left(\begin{tabular}{l|lllllllllllllllllllllllllllll}
		& $E_1$ & $E_2$ & $E_3$ & $E_4$ & $E_5$  & $E_6$ & $E_7$ & $E_8$ & $E_9$ & $E_{10}$ & $E_{11}$ & $E_{12}$ & $E_{13}$ & $E_{14}$  & $E_{15}$ & $E_{16}$ & $E_{17}$ & $E_{18}$ & $E_{19}$ & $E_{20}$ & $E_{21}$ & $E_{22}$ & $E_{23}$  & $E_{24}$ & $E_{25}$ & $E_{26}$  & $E_{27}$   \\
		\hline
		$T_1$     & 2 & 3 & 4 & 5 & 6 & 7 & 0 & 0 & 0 & 0 & 0 & 0 & 0 & 0 & 0 & 0 & 0 & 0 & 0 & 0 & 0 & 0 & 0 & 0 & 0 & 0 & 0  \\
		$T_2$     & 2 & 0 & 0 & 0 & 0 & 0 & 3 & 4 & 5 & 6 & 7 & 0 & 0 & 0 & 0 & 0 & 0 & 0 & 0 & 0 & 0 & 0 & 0 & 0 & 0 & 0 & 0  \\
		$T_3$     & 0 & 2 & 0 & 0 & 0 & 0 & 3 & 0 & 0 & 0 & 0 & 4 & 5 & 6 & 7 & 0 & 0 & 0 & 0 & 0 & 0 & 0 & 0 & 0 & 0 & 0 & 0  \\
		$T_4$     & 0 & 0 & 2 & 0 & 0 & 0 & 0 & 0 & 0 & 0 & 0 & 3 & 0 & 0 & 0 & 4 & 5 & 6 & 7 & 0 & 0 & 0 & 0 & 0 & 0 & 0 & 0  \\
		$T_5$     & 0 & 0 & 0 & 2 & 0 & 0 & 0 & 3 & 0 & 0 & 0 & 0 & 0 & 0 & 0 & 4 & 0 & 0 & 0 & 5 & 6 & 7 & 0 & 0 & 0 & 0 & 0  \\
		$T_6$     & 0 & 0 & 0 & 0 & 0 & 0 & 0 & 0 & 2 & 0 & 0 & 0 & 3 & 0 & 0 & 0 & 4 & 0 & 0 & 5 & 0 & 0 & 6 & 7 & 0 & 0 & 0  \\
		$T_7$     & 0 & 0 & 0 & 0 & 2 & 0 & 0 & 0 & 0 & 3 & 0 & 0 & 0 & 0 & 0 & 0 & 0 & 4 & 0 & 0 & 0 & 0 & 5 & 0 & 6 & 7 & 0  \\
		$T_8$     & 0 & 0 & 0 & 0 & 0 & 0 & 0 & 0 & 0 & 0 & 2 & 0 & 0 & 3 & 0 & 0 & 0 & 0 & 4 & 0 & 5 & 0 & 0 & 0 & 6 & 0 & 7  \\
		$T_9$     & 0 & 0 & 0 & 0 & 0 & 2 & 0 & 0 & 0 & 0 & 0 & 0 & 0 & 0 & 3 & 0 & 0 & 0 & 0 & 0 & 0 & 4 & 0 & 5 & 0 & 6 & 7  \\
		
	\end{tabular}\right),}
\label{Eq:smblbq}
\ee

\subsection{$2d$ kagome lattice}
\label{appx:SMkagome}

\begin{figure}[h]
	\centerline{\includegraphics[width=13cm]{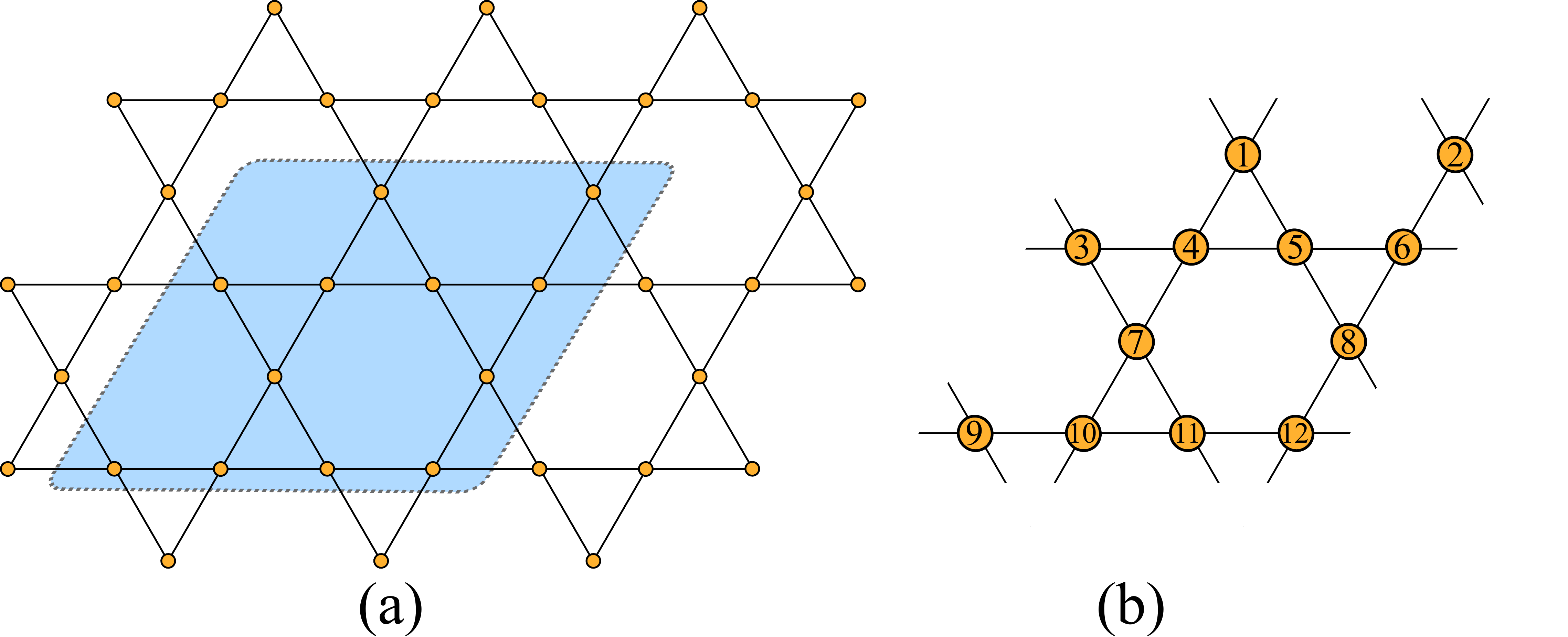}}
	\caption{(Color online) (a) The infinite $2d$ kagome lattice with a 12-site unit-cell (blue region). (b) Labeling of vertices (graph nodes) in the unit-cell.}
	\label{Fig:kagome}
\end{figure}

Eq.\eqref{Eq:smkagome} corresponds to the SM of an infinite $2d$ kagome lattice with a 12-site unit-cell (See Fig.~\ref{Fig:kagome}).

\be
SM_{\rm kagome}={\small \left(\begin{tabular}{l|llllllllllllllllllllllllll}
		& $E_1$ & $E_2$ & $E_3$ & $E_4$ & $E_5$  & $E_6$ & $E_7$ & $E_8$ & $E_9$ & $E_{10}$ & $E_{11}$ & $E_{12}$ & $E_{13}$ & $E_{14}$  & $E_{15}$ & $E_{16}$ & $E_{17}$ & $E_{18}$ & $E_{19}$ & $E_{20}$ & $E_{21}$ & $E_{22}$ & $E_{23}$  & $E_{24}$  \\
		\hline
		$T_1$     & 2 & 3 & 4 & 5 & 0 & 0 & 0 & 0 & 0 & 0 & 0 & 0 & 0 & 0 & 0 & 0 & 0 & 0 & 0 & 0 & 0 & 0 & 0 & 0  \\
		$T_2$     & 0 & 0 & 0 & 0 & 2 & 3 & 4 & 5 & 0 & 0 & 0 & 0 & 0 & 0 & 0 & 0 & 0 & 0 & 0 & 0 & 0 & 0 & 0 & 0  \\
		$T_3$     & 0 & 0 & 0 & 0 & 2 & 0 & 0 & 0 & 3 & 4 & 5 & 0 & 0 & 0 & 0 & 0 & 0 & 0 & 0 & 0 & 0 & 0 & 0 & 0  \\
		$T_4$     & 2 & 0 & 0 & 0 & 0 & 0 & 0 & 0 & 3 & 0 & 0 & 4 & 5 & 0 & 0 & 0 & 0 & 0 & 0 & 0 & 0 & 0 & 0 & 0  \\
		$T_5$     & 0 & 2 & 0 & 0 & 0 & 0 & 0 & 0 & 0 & 0 & 0 & 3 & 0 & 4 & 5 & 0 & 0 & 0 & 0 & 0 & 0 & 0 & 0 & 0  \\
		$T_6$     & 0 & 0 & 0 & 0 & 0 & 2 & 0 & 0 & 0 & 3 & 0 & 0 & 0 & 4 & 0 & 5 & 0 & 0 & 0 & 0 & 0 & 0 & 0 & 0  \\
		$T_7$     & 0 & 0 & 0 & 0 & 0 & 0 & 0 & 0 & 0 & 0 & 2 & 0 & 3 & 0 & 0 & 0 & 4 & 5 & 0 & 0 & 0 & 0 & 0 & 0  \\
		$T_8$     & 0 & 0 & 0 & 0 & 0 & 0 & 0 & 0 & 0 & 0 & 0 & 0 & 0 & 0 & 2 & 3 & 0 & 0 & 4 & 5 & 0 & 0 & 0 & 0  \\
		$T_9$     & 0 & 0 & 2 & 0 & 0 & 0 & 0 & 0 & 0 & 0 & 0 & 0 & 0 & 0 & 0 & 0 & 0 & 0 & 3 & 0 & 4 & 5 & 0 & 0  \\
		$T_{10}$  & 0 & 0 & 0 & 2 & 0 & 0 & 0 & 0 & 0 & 0 & 0 & 0 & 0 & 0 & 0 & 0 & 3 & 0 & 0 & 0 & 4 & 0 & 5 & 0  \\
		$T_{11}$  & 0 & 0 & 0 & 0 & 0 & 0 & 2 & 0 & 0 & 0 & 0 & 0 & 0 & 0 & 0 & 0 & 0 & 3 & 0 & 0 & 0 & 0 & 4 & 5  \\
		$T_{12}$  & 0 & 0 & 0 & 0 & 0 & 0 & 0 & 2 & 0 & 0 & 0 & 0 & 0 & 0 & 0 & 0 & 0 & 0 & 0 & 3 & 0 & 4 & 0 & 5  \\	
		
	\end{tabular}\right).}
\label{Eq:smkagome}
\ee

\subsection{$3d$ cubic lattice}
\label{appx:SMcubic}

\begin{figure}[h]
	\centerline{\includegraphics[width=4cm]{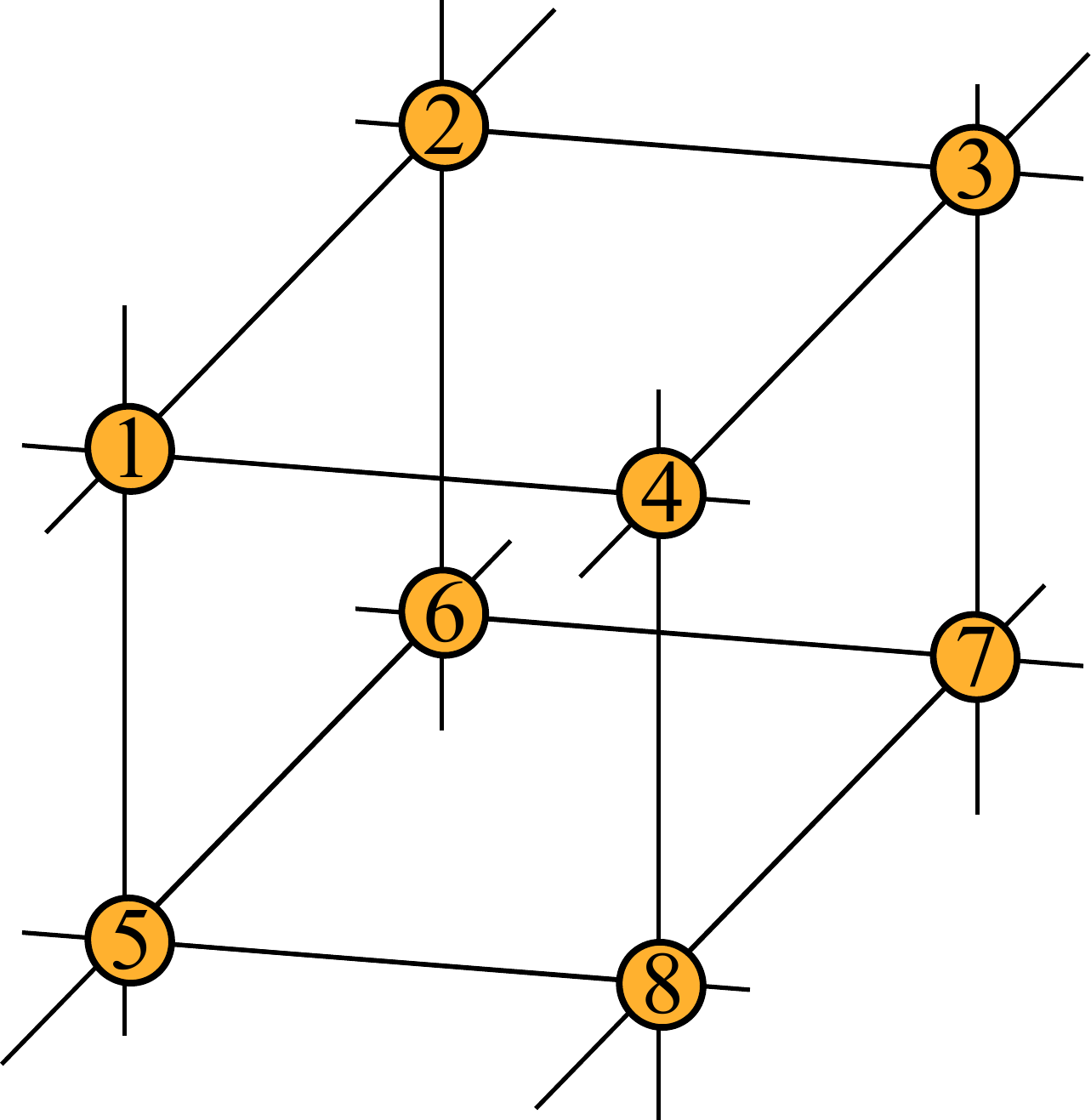}}
	\caption{(Color online) The infinite $3d$ cubic lattice with a 8-site unit-cell. The numbers at vertices label the graph nodes in the unit-cell.}
	\label{Fig:cubic}
\end{figure}

Eq.\eqref{Eq:smcubic} corresponds to the SM of an $3d$ cubic lattice with a 8-site unit-cell. (See Fig.~\ref{Fig:cubic}).

\be
SM_{\rm cube}={\small \left(\begin{tabular}{l|llllllllllllllllllllllllll}
		& $E_1$ & $E_2$ & $E_3$ & $E_4$ & $E_5$  & $E_6$ & $E_7$ & $E_8$ & $E_9$ & $E_{10}$ & $E_{11}$ & $E_{12}$ & $E_{13}$ & $E_{14}$  & $E_{15}$ & $E_{16}$ & $E_{17}$ & $E_{18}$ & $E_{19}$ & $E_{20}$ & $E_{21}$ & $E_{22}$ & $E_{23}$  & $E_{24}$  \\
		\hline
		$T_1$     & 2 & 3 & 4 & 5 & 6 & 7 & 0 & 0 & 0 & 0 & 0 & 0 & 0 & 0 & 0 & 0 & 0 & 0 & 0 & 0 & 0 & 0 & 0 & 0  \\
		$T_2$     & 2 & 3 & 0 & 0 & 0 & 0 & 4 & 5 & 6 & 7 & 0 & 0 & 0 & 0 & 0 & 0 & 0 & 0 & 0 & 0 & 0 & 0 & 0 & 0  \\
		$T_3$     & 0 & 0 & 0 & 0 & 0 & 0 & 2 & 3 & 0 & 0 & 4 & 5 & 6 & 7 & 0 & 0 & 0 & 0 & 0 & 0 & 0 & 0 & 0 & 0  \\
		$T_4$     & 0 & 0 & 2 & 3 & 0 & 0 & 0 & 0 & 0 & 0 & 4 & 5 & 0 & 0 & 6 & 7 & 0 & 0 & 0 & 0 & 0 & 0 & 0 & 0  \\
		$T_5$     & 0 & 0 & 0 & 0 & 2 & 3 & 0 & 0 & 0 & 0 & 0 & 0 & 0 & 0 & 0 & 0 & 4 & 5 & 6 & 7 & 0 & 0 & 0 & 0  \\
		$T_6$     & 0 & 0 & 0 & 0 & 0 & 0 & 0 & 0 & 2 & 3 & 0 & 0 & 0 & 0 & 0 & 0 & 4 & 5 & 0 & 0 & 6 & 7 & 0 & 0  \\
		$T_7$     & 0 & 0 & 0 & 0 & 0 & 0 & 0 & 0 & 0 & 0 & 0 & 0 & 2 & 3 & 0 & 0 & 0 & 0 & 0 & 0 & 4 & 5 & 6 & 7  \\
		$T_8$     & 0 & 0 & 0 & 0 & 0 & 0 & 0 & 0 & 0 & 0 & 0 & 0 & 0 & 0 & 2 & 3 & 0 & 0 & 4 & 5 & 0 & 0 & 6 & 7  \\
		
	\end{tabular}\right).}
\label{Eq:smcubic}
\ee

\subsection{$3d$ pyrochlore lattice}
\label{appx:SMpyrochlore}

\begin{figure}[h]
	\centerline{\includegraphics[width=12cm]{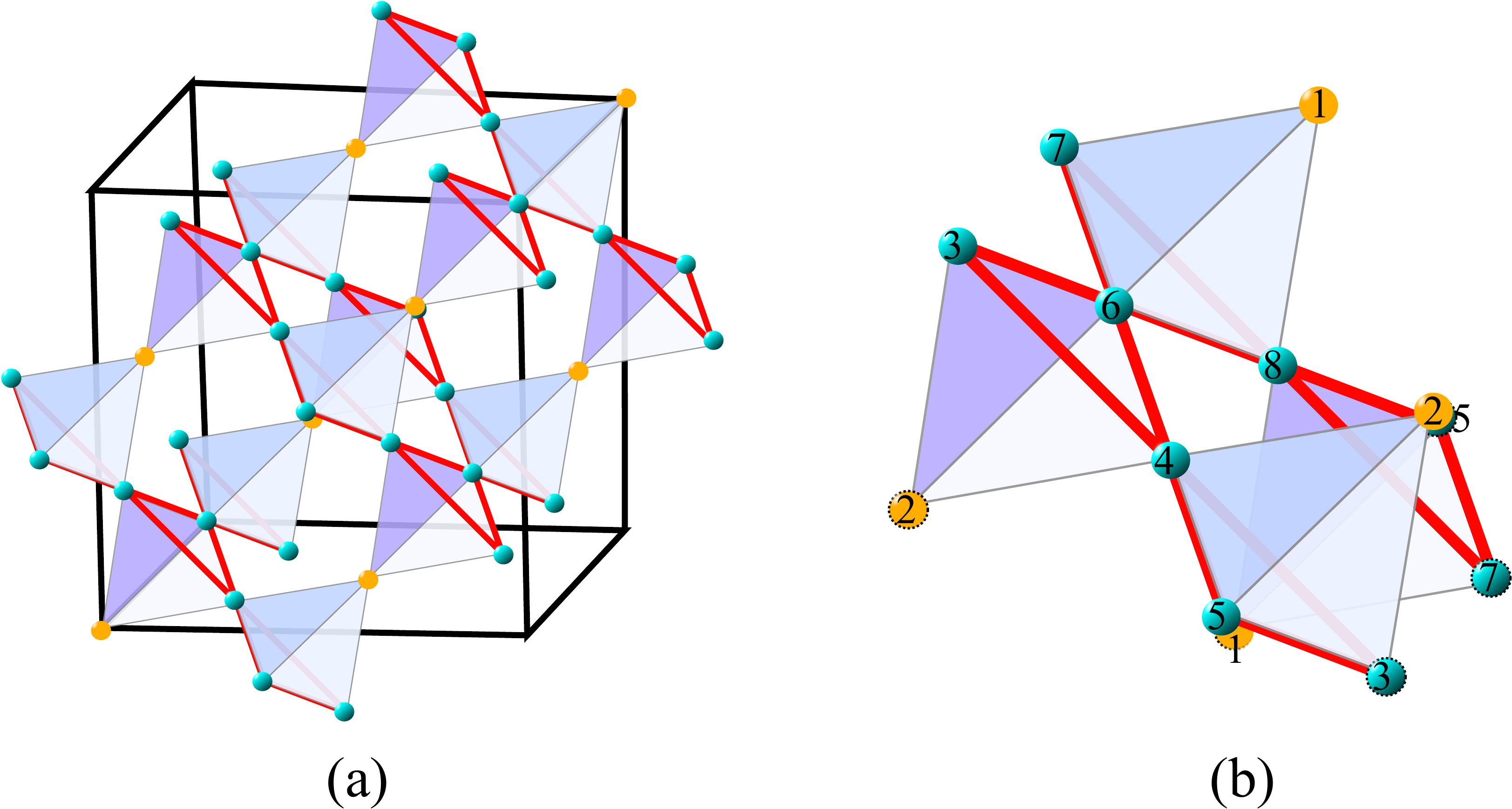}}
	\caption{(Color online) (a) The infinite $3d$ pyrochlore lattice composed of up and down tetrahedrons. (b) The eight-site unit-cell of the pyrochlore lattice. The numbers represent the labeling of vertices (graph nodes) in the unit-cell.}
	\label{Fig:pyrochlor}
\end{figure}

Eq.\eqref{Eq:smpyrochlore} corresponds to the SM of an $3d$ pyrochlore lattice with a 8-site unit-cell. (See Fig.~\ref{Fig:pyrochlor}).

\be
SM_{\rm pyro}={\small \left(\begin{tabular}{l|llllllllllllllllllllllllll}
		& $E_1$ & $E_2$ & $E_3$ & $E_4$ & $E_5$  & $E_6$ & $E_7$ & $E_8$ & $E_9$ & $E_{10}$ & $E_{11}$ & $E_{12}$ & $E_{13}$ & $E_{14}$  & $E_{15}$ & $E_{16}$ & $E_{17}$ & $E_{18}$ & $E_{19}$ & $E_{20}$ & $E_{21}$ & $E_{22}$ & $E_{23}$  & $E_{24}$  \\
		\hline
		$T_1$     & 2 & 3 & 4 & 5 & 6 & 7 & 0 & 0 & 0 & 0 & 0 & 0 & 0 & 0 & 0 & 0 & 0 & 0 & 0 & 0 & 0 & 0 & 0 & 0  \\
		$T_2$     & 0 & 0 & 0 & 0 & 0 & 0 & 2 & 3 & 4 & 5 & 6 & 7 & 0 & 0 & 0 & 0 & 0 & 0 & 0 & 0 & 0 & 0 & 0 & 0  \\
		$T_3$     & 0 & 0 & 0 & 0 & 0 & 0 & 2 & 3 & 0 & 0 & 0 & 0 & 4 & 5 & 6 & 7 & 0 & 0 & 0 & 0 & 0 & 0 & 0 & 0  \\
		$T_4$     & 0 & 0 & 0 & 0 & 0 & 0 & 0 & 0 & 2 & 3 & 0 & 0 & 4 & 5 & 0 & 0 & 6 & 7 & 0 & 0 & 0 & 0 & 0 & 0  \\
		$T_5$     & 2 & 0 & 0 & 0 & 0 & 0 & 0 & 0 & 0 & 0 & 3 & 0 & 0 & 0 & 4 & 0 & 5 & 0 & 6 & 7 & 0 & 0 & 0 & 0  \\
		$T_6$     & 0 & 2 & 0 & 0 & 0 & 0 & 0 & 0 & 0 & 0 & 0 & 3 & 0 & 0 & 0 & 4 & 0 & 5 & 0 & 0 & 6 & 7 & 0 & 0  \\
		$T_7$     & 0 & 0 & 2 & 3 & 0 & 0 & 0 & 0 & 0 & 0 & 0 & 0 & 0 & 0 & 0 & 0 & 0 & 0 & 4 & 0 & 5 & 0 & 6 & 7  \\
		$T_8$     & 0 & 0 & 0 & 0 & 2 & 3 & 0 & 0 & 0 & 0 & 0 & 0 & 0 & 0 & 0 & 0 & 0 & 0 & 0 & 4 & 0 & 5 & 6 & 7  \\
		
	\end{tabular}\right).}
\label{Eq:smpyrochlore}
\ee

\subsection{$3d$ hyperhoneycomb lattice}
\label{appx:SMhyperhoneycomb}

\begin{figure}[h]
	\centerline{\includegraphics[width=14cm]{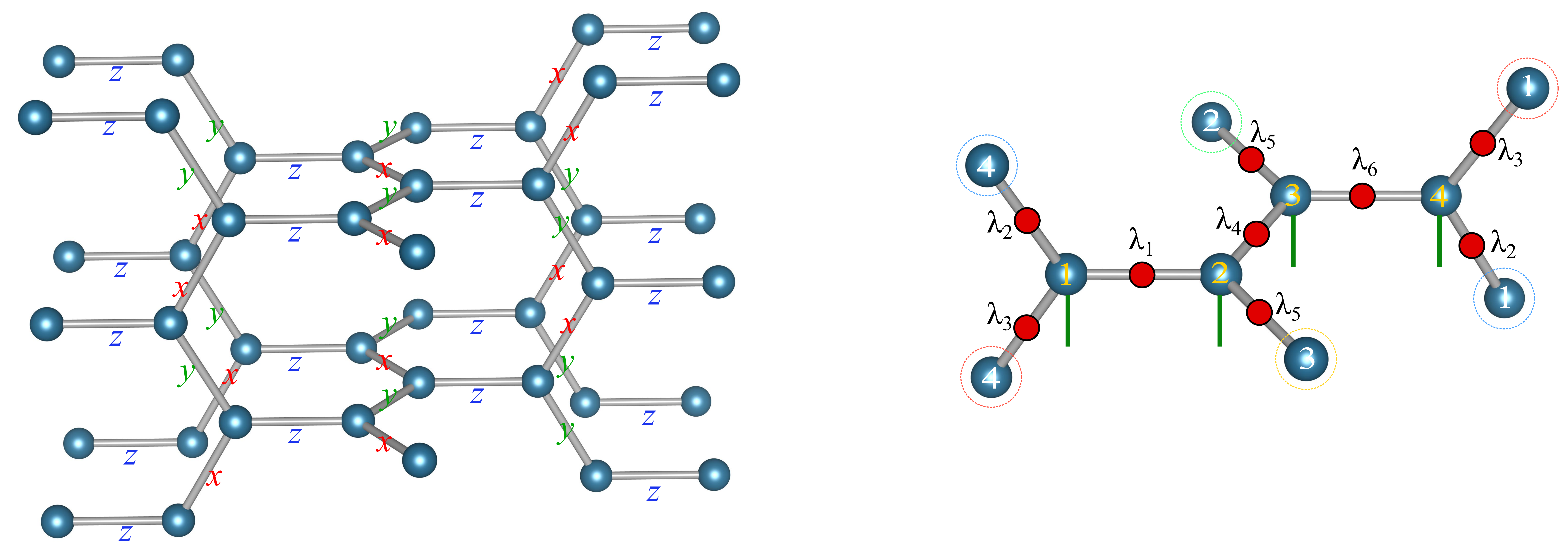}}
	\caption{(Color online) (a) The infinite $3d$ hyperhoneycomb lattice. (b) The four-site unit-cell of the hyperhoneycomb lattice. The numbers represent the labeling of vertices (graph nodes) in the unit-cell.}
	\label{Fig:hyperhoneycomb}
\end{figure}

Eq.\eqref{Eq:smhyperhoney} corresponds to the SM of an $3d$ hyperhoneycomb lattice with a 4-site unit-cell. (See Fig.~\ref{Fig:hyperhoneycomb}).

\be
\left(\begin{tabular}{l|lllllllll}
	& $E_1$ & $E_2$ & $E_3$ & $E_4$ & $E_5$  & $E_6$ \\
	\hline
	$T_1$     & 2 & 3 & 4 & 0 & 0 & 0 \\
	$T_2$     & 2 & 0 & 0 & 3 & 4 & 0 \\
	$T_3$     & 0 & 0 & 0 & 2 & 3 & 4 \\
	$T_4$     & 0 & 2 & 3 & 0 & 0 & 4 \\
\end{tabular}\right).
\label{Eq:smhyperhoney}
\ee

\section{Gauge-Fixing for gPEPS}
\label{appx:gauge}

\begin{figure}[h]
	\centerline{\includegraphics[width=14cm]{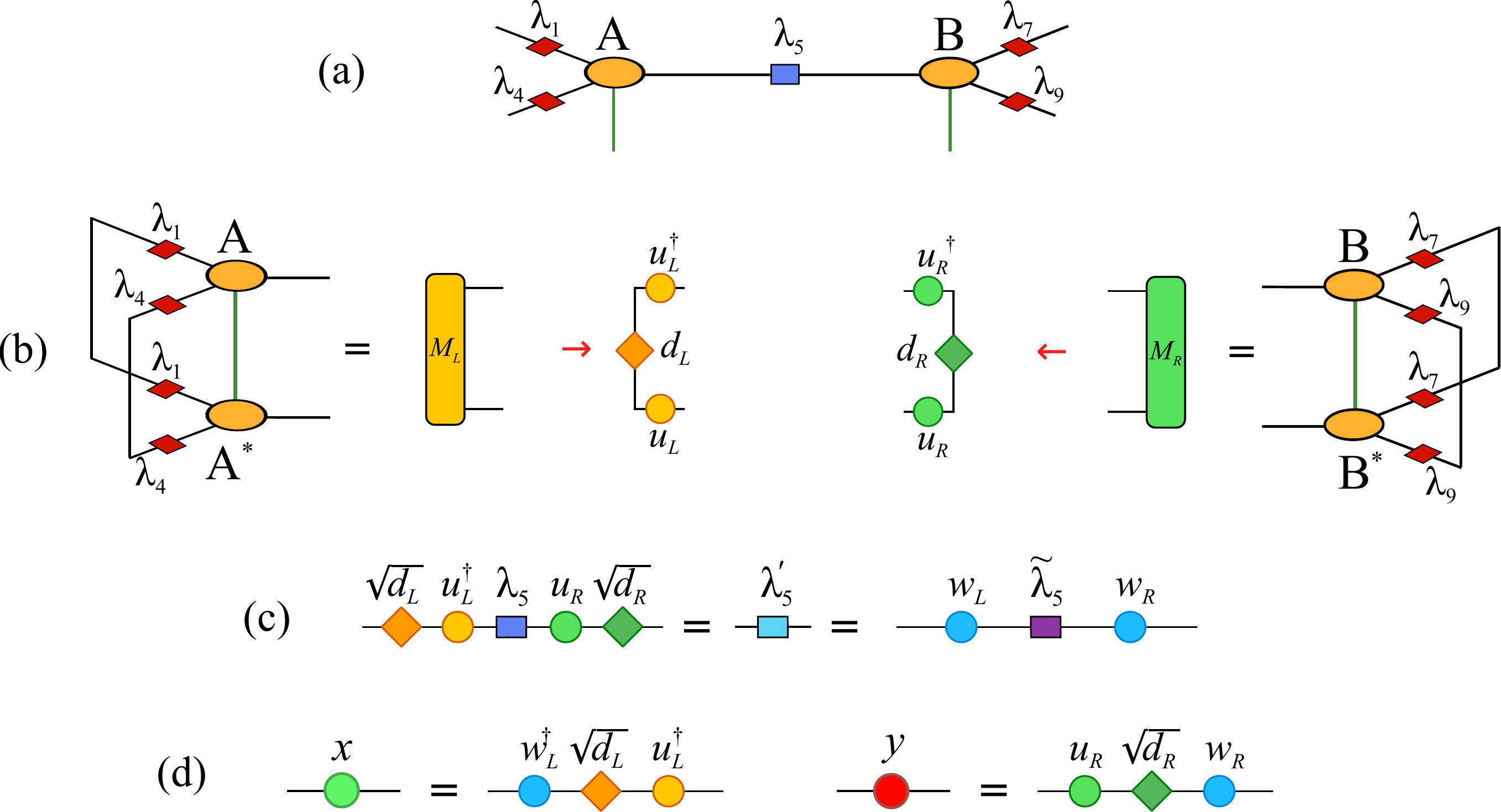}}
	\caption{(Color online) (a) Local state $\ket{\psi}$ composed of tensors $A$, $B$ and their relevant $\lambda$ matrices as effective mean-field environment. (b) Boundary matrices $M_L$ and $M_R$ and their eigendecompositions. (c) Definition of modified bond matrix $\lambda'$, which is then decomposed via the SVD. (d) Definition of the gauge change matrices $x$ and $y$ that transform the initial state $\ket{\psi}$ to its gauge related state $\ket{\tilde{\psi}}$.}
	\label{Fig:super-ortho}
\end{figure}

In this section, we show how to locally fix the gauge degrees of freedom on the virtual bonds of the gPEPS TN. This can substantially improve the algorithm by stabilizing the ITE optimization and results in faster convergence of the ITE iteration and more accurate estimation of expectation values and correlators. To this end, we first introduce the boundary matrices for each link of the TN: consider a virtual bond of a TN shared between tensor $A, B$ and their corresponding $\lambda$ matrices, such as the one shown in Fig.~\ref{Fig:super-ortho}-(a), the left and right boundary matrices are defined as (see also Fig.~\ref{Fig:super-ortho}-(b))
\bea
\left( {{M _L}} \right)_{i'}^i & = \sum\limits_{p,j,k,p',j',k'} A_{i,j,k}^p \bar{A}_{i',j',k'}^{p'} {\lambda _{jj'}^2}{\lambda _{kk'}^2} \hfill \nonumber\\
\left( {{M _R}} \right)_{j'}^j & = \sum\limits_{p,i,k,p',i',k'} B_{i,j,k}^p \bar{B}_{i',j',k'}^{p'} {\lambda _{ii'}^2}{\lambda _{kk'}^2} \hfill 
\label{Eq:MLMR}.
\eea

We choose the gauge degrees of freedom such that a Schmidt form is imposed on all virtual degrees of freedom on the TN network. this involves choosing the gauge such that (i) the $M_L$ and $M_R$ boundary matrices represent an orthonormal basis i.e. $M^i_{i'}=\lambda_i^2 \delta_{i,i'}$,  and (ii) the bond matrices $\lambda$ are diagonal, normalized and positive, $\lambda_{i,j}=\delta_{i,j} s_i$ with $s_i$ the Schmidt coefficients, which are ordered $s_i \ge s_{i+1}$. A {\it canonical form} for the tensor network is defined by requiring that every virtual bond is in Schmidt form \cite{Ran2012,Evenbly2018}.

\begin{figure}[h]
	\centerline{\includegraphics[width=10cm]{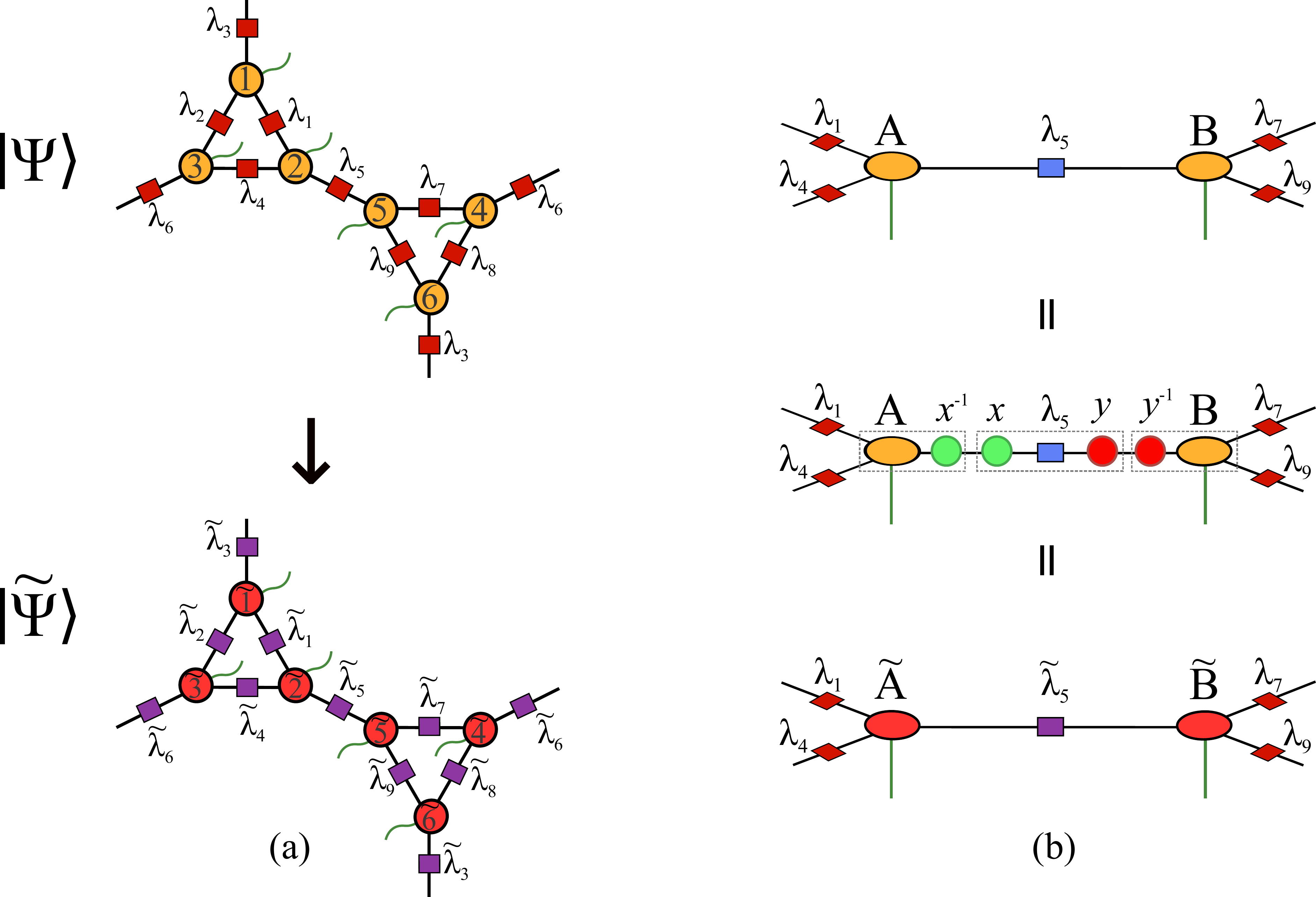}}
	\caption{(Color online) (a)  Gauge related initial and final states. (b) A change of gauge, which leaves the state $\ket{\psi}$ invariant,
		is enacted on the index between $A$ and $B$ via matrices $x$ and
		$y$ together with their inverses.}
	\label{Fig:gauge-fix}
\end{figure}

We now present a method to fix the gauge degrees of freedom on any virtual link of a given network. Note that under change of the gauge all local tensors associated to a link are altered i.e., $A\rightarrow\tilde{A}$, $B\rightarrow\tilde{B}$ and the shared lambda matrix $\lambda\rightarrow\tilde{\lambda}$. Correspondingly, after applying the gauge-fixing to all virtual bonds of the TN the iPEPS wave-function is altered as well (see Fig.~\ref{Fig:gauge-fix}-(a)).

In order to identifying the gauge change matrices $x$ and $y$ (and their inverses) we first calculate the boundary matrices $M_L$ and $M_R$ and then diagonalize them such that
\bea
{M_L} &= {u_L} {{d_L}} u_L^\dag, \nonumber \\
{M_R} &= {u_R} {{d_R}} u_R^\dag, 
\label{Eq:MLReigendec}
\eea
see Fig.~\ref{Fig:super-ortho}-(b), with unitary matrices $u_L$ , $u_R$ and real diagonal matrices $d_L$ , $d_R$. Notice that, due to the positivity of the boundary matrices $M_L$ and $M_R$, it follows  that $d_L$ and $d_R$ are non-negative, thus possess real roots $\sqrt{d_L}$ and $\sqrt{d_R}$ . We now use these to
transform the bond matrix $\lambda$,
\be
\lambda ' \equiv \sqrt {{d_L}} u_L^\dag \lambda \  u_R \sqrt{{d_R}}, 
\label{Eq:lambdaprim}
\ee
and take the singular value decomposition to obtain
\be
\lambda ' = {w_L}\tilde \lambda w_R^\dag 
\label{Eq:lambdapsvd}
\ee
for unitary $w_L$, $w_R$ and positive diagonal $\tilde \lambda$. The gauge change matrices $x$ and  $y$ are now defined as
\bea
x &\equiv w_L^\dag \sqrt {{d_L}} u_L^\dag, \nonumber\\
y &\equiv u_R \sqrt {{d_R}} w_R. 
\label{Eq:gaugexy}
\eea
This process is further depicted in Fig.~\ref{Fig:super-ortho}-(c-d). One should note that $M_L=x^\dagger x$ and $M_R=yy^\dagger$ . Under this choice of gauge the new
bond matrix is simply the $\tilde{\lambda}$ from Eq.\eqref{Eq:lambdapsvd} or equivalently 
\be
\tilde{\lambda}=x\lambda y, 
\ee
which is positive and diagonal by construction. Furthermore the new left and right tensor read
\bea
\tilde{A} &= A x^{-1}, \nonumber\\
\tilde{B} &= y^{-1} B,
\label{Eq:newAB}
\eea
see also Fig.~\ref{Fig:gauge-fix}-(b). Once the gauge is fixed on all virtual legs of the $A$, $B$ tensors the Schmidt form or orthonormality are satisfied when the eigenvalues in $d_L$ , $d_R$ are uniformly distributed i.e., all the diagonal elements are equals to $1$ and $M^i_{i'}-\lambda_i^2 \delta_{i,i'}=0$ for both left and right boundary matrices. 

in Ref~.\cite{Ran2012}. this process of gauge-fixing is alternatively dubbed as super-orthogonality and is also equivalent as doing high-order SVD on local tensors. In order to bring all of the tensors in a TN into a super-orthogonal form, one can iteratively do the above process or rather incorporate it into the simple-update optimization and fix the gauge on all tensors before every step of ITE. We refer the interested reader for detailed discussion on this subject to Ref.~\cite{Ran2012,Evenbly2018}.

\end{document}